\documentclass[a4paper,fleqn]{cas-sc}

\usepackage[numbers,square,sort&compress]{natbib}
\usepackage{graphicx}%
\usepackage{multirow}%
\usepackage{amsmath,amssymb,amsfonts}%
\usepackage{amsthm}%
\usepackage{mathrsfs}%
\usepackage[title]{appendix}%
\usepackage{xcolor}%
\usepackage{textcomp}%
\usepackage{manyfoot}%
\usepackage{booktabs}%
\usepackage{algorithm}%
\usepackage{algorithmicx}%
\usepackage{algpseudocode}%
\usepackage{listings}%
\usepackage{makecell}%

\usepackage{hyperref}   
\hypersetup{
	colorlinks=true,
	linkcolor=blue,
	citecolor=blue,
	urlcolor=blue,
}

\usepackage{mathtools}

\usepackage{lipsum}
\usepackage{diagbox}
\usepackage{indentfirst}

\usepackage{array}
\usepackage{tabularx}
\usepackage{kotex}
\usepackage{float}%
\usepackage{capt-of}

\newcommand{\cnum}[1]{\text{\textcircled{\scriptsize #1}}}
\newcommand{\dia}{\varnothing}
\newcommand{\fid}[1]{\textsf{#1}}

\def\tsc#1{\csdef{#1}{\textsc{\lowercase{#1}}\xspace}}
\tsc{WGM}
\tsc{QE}
\tsc{EP}
\tsc{PMS}
\tsc{BEC}
\tsc{DE}

\begin{document}
\let\WriteBookmarks\relax
\shorttitle{Design-Intent-Aware Reverse Parametrization}
\shortauthors{Y. Lee et~al.}

\title [mode = title]{From Static Models to Editable CAD: A Design-Intent-Aware Framework for Reverse Parametrization}

\author[1]{Yurim Lee}
\ead{yurim.lee@kaist.ac.kr}
\credit{Conceptualization of this study, Data curation, Writing - Original draft preparation}
\author[2]{Sumin Lee}
\cormark[1]
\ead{sumlee@jbnu.ac.kr}
\credit{Methodology, Data curation, Writing - Original draft preparation}
\author[4,3]{Namwoo Kang}
\cormark[1]   
\ead{nwkang@kaist.ac.kr}
\credit{Conceptualization of this study, Data curation, Writing - Original draft preparation }

\affiliation[1]{Department of Mechanical Engineering, Korea Advanced Institute of Science and Technology (KAIST), 291 Daehak-ro, Yuseong-gu, Daejeon 34141, Republic of Korea}

\affiliation[2]{Department of Mechanical System Engineering, Jeonbuk National University (JBNU), 567 Beakje-daero, duckjin-gu Jeonbuk 54896, Republic of Korea}

\affiliation[3]{Narnia Labs, 193 Munji-ro, Yuseong-gu, Daejeon 34051, Republic of Korea}

\affiliation[4]{Department of Mechanical Engineering, Korea Advanced Institute of Science and Technology (KAIST), 291 Daehak-ro, Yuseong-gu, Daejeon 34141, Republic of Korea}

\cortext[cor1]{Corresponding author}


\begin{abstract}
Neutral CAD formats such as STEP typically preserve geometry but not feature histories or parameter dependencies. Thus, they are inherently limited for iterative design that requires modifications, a critical limitation for design optimization that relies on iterative processes.
This study presents a design-intent-aware reverse-parametrization framework that converts history-free STEP parts into editable Grasshopper definitions and represents their assembly relations. Without task-specific training, the framework combines kernel-measured features, geometric relations, a compressed attributed adjacency graph, and multiview renderings. An LLM assigns dimensions independent, fixed, or dependent roles, confirms candidate joints, and interprets verification reports. Kernel- and rule-based modules perform measurement, parameter binding, geometry construction, and verification. Evaluation distinguishes default-shape preservation, design-intent reflection, and edit validity, with the latter examined through assembly-level editing examples involving checks for geometry regeneration, joint alignment, interference, and mechanism closure.
As an example, five components of a mechanical mechanism were used, and across 25 runs, the mean global and analytic-feature-region intersection over unions (IoUs) were 0.984 and 0.983, respectively; all runs passed the combined shape-preservation criterion. 
In a separate request experiment, the requested dimension was selected as an independent variable in 19 of 20 runs. Product-level examples show coordinated changes in part geometry and assembly design while preserving assembly relations and avoiding inter-part interference. These results demonstrate editable reconstruction of history-free CAD models under the evaluated geometric and assembly constraints, indicating potential for subsequent design optimization.

\end{abstract}

\begin{graphicalabstract}
\centering
\includegraphics[
    width=\textwidth,
    height=0.35\textheight,
    keepaspectratio
]{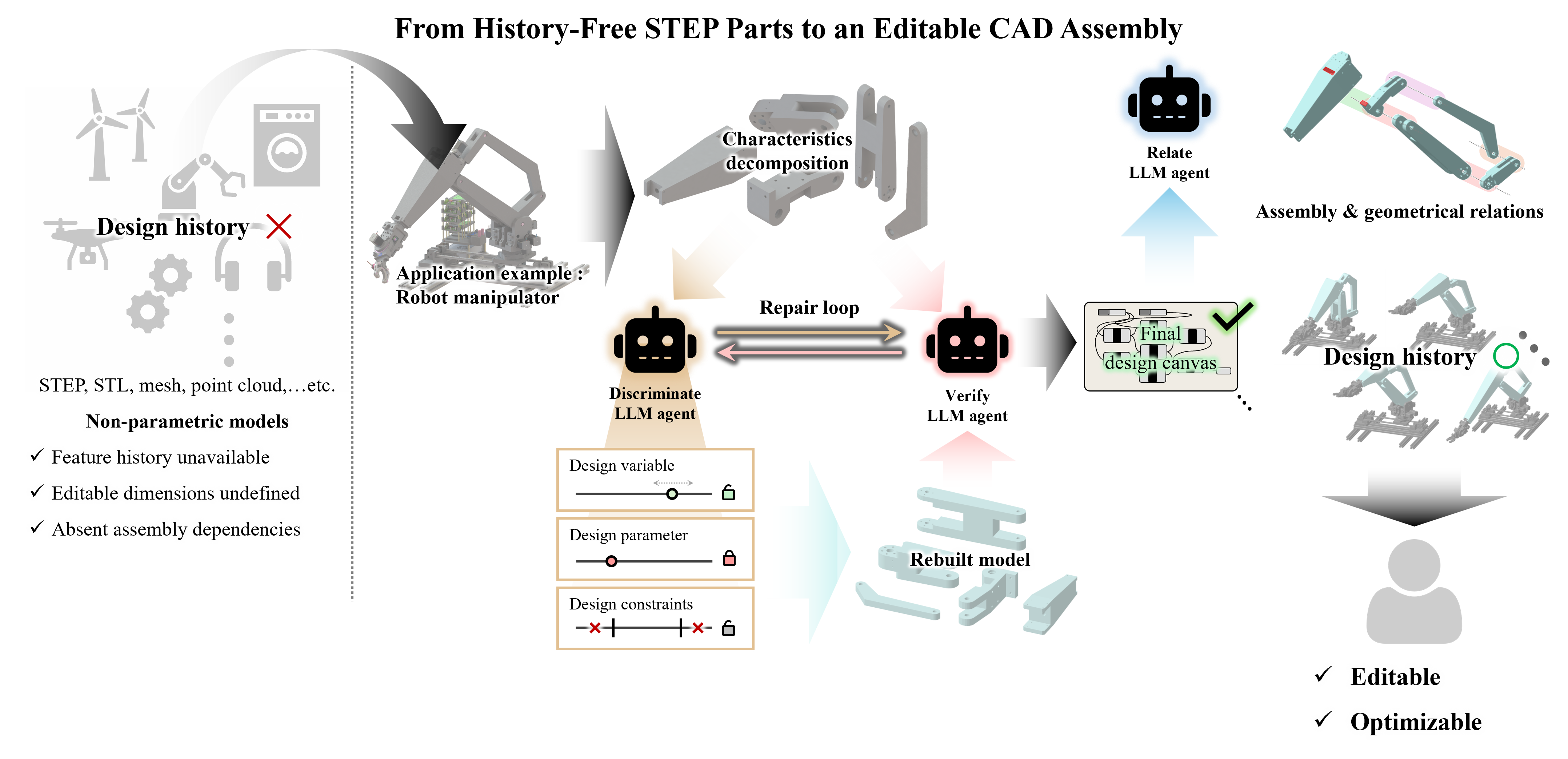}

\end{graphicalabstract}

\begin{highlights}
\item History-free STEP parts are reconstructed into editable parametric CAD models.
\item Design intent is reflected through independent and dependent design variables.
\item Assembly relations are preserved during coordinated parametric editing.
\item Shared parameters enable coordinated changes in component geometry and assembly.
\end{highlights}


\maketitle

\section{Introduction}

Computer-aided design (CAD) models serve as the standard medium of modern product development, supporting geometric definition, engineering analysis, process planning, and manufacturing. Design decisions are represented through the dimensions and geometry of these models. Product design in industrial practice is inherently iterative: changes in manufacturing conditions, such as tolerances and tooling processes, and in product requirements repeatedly necessitate dimensional modifications. Parametric CAD models accommodate such changes by defining principal dimensions as parameters and preserving the reference relationships among modeling features. However, the robustness and reusability of parametric CAD models depend not only on the availability of parameters but also on the organization of feature dependencies and modeling strategies \citep{camba2016parametric}.

In contrast, modifying dimensions systematically in a non-parametric model can be difficult. Relevant faces may need to be edited individually, and related features may not update together. This loss of associativity prevents design variables that encode the designer’s intent from being modified systematically, thereby making the underlying design history difficult to trace and substantially hindering iterative design. As changes accumulate, the required work may approach complete remodeling. A non-parametric model can still be used to inspect geometry or support manufacturing and analysis. However, the lack of structured editability makes such models difficult to incorporate into iterative design cycles in which products are progressively refined through dimensional changes. Despite this limitation, non-parametric models remain widespread in practice. In particular, product geometry is frequently handed off in non-parametric form when development moves from industrial styling to detailed engineering design. This commonly occurs when physical mock-ups or existing products are digitized through 3D scanning, when three-dimensional forms are reconstructed from 2D sketches or images, and when legacy designs are recovered through reverse-engineering workflows. Although these processes can reproduce or generate the intended geometry, they generally yield mesh or static B-rep representations without the design variables, feature dependencies, and construction history required for subsequent engineering modification. A similar loss of editability can occur during cross-system CAD data exchange. Although standardized approaches to transferring feature histories, parameters, and constraints have been investigated, geometry-oriented data exchange may provide boundary representations without the original modeling history and parameter dependencies \cite{pratt2005towards,kim2008standardized}. Consequently, a gap exists between models that must be used and models that can be effectively used. This study addresses this gap through reverse parametrization, namely, transforming a set of non-parametric part models into an editable representation of the product they form.

Moreover, the dimensions of a part acquire much of their meaning from the parts it is assembled with. Many of them exist only to fit its counterpart, and a change to one of them is in practice a change to the pair. A part made parametric in isolation therefore remains only partly reusable: its dimensions can be edited, but the relations that constrain those edits are not represented, so nothing in the model indicates when an edit has made the product impossible to assemble. Reverse parametrization that stops at the part boundary recovers dimensions without recovering the relations that give them meaning.

The key criterion that determines whether a model can be reused is not whether it is nominally parametric, or merely whether parameters exist, but whether it is editable. Non-parametric models are excluded from iterative design workflows because they cannot be readily modified. By the same criterion, a model that contains parameters but does not permit meaningful modification remains effectively equivalent to a non-parametric model.
Once editability is adopted as the criterion, the question becomes what should be made editable. Design intent broadly encompasses the functional roles of features, dependencies among dimensions, and the degrees of freedom intended to remain changeable \cite{otey2018revisiting}. Within this broader concept, the present study defines the operational scope of design intent for reverse parametrization as determining which dimensions should remain independently editable, which should be fixed, and which should depend on other reference dimensions--within a part, and across the parts of a product, where a dependency becomes a shared variable or a measured coupling between two parts. What industrial users ultimately require is not merely geometric reproduction, but the recovery of this design intent. However, this information is generally not explicitly recorded in a history-free model. It resides in the designer's reasoning or, at best, is maintained in checklists and design documents rather than in a machine-readable form connected to the model. In current practice, an experienced designer must therefore inspect a STEP model, infer its intent, and manually remodel it. This is a costly task that must be repeated for each part.

Research on reconstructing parametric representations from history-free models has developed along several directions, including constructive solid geometry (CSG) recovery, B-rep reconstruction, CAD command-sequence generation, and large language model (LLM)-based code generation. CSG-based methods provide editable structures through Boolean combinations of analytic primitives \cite{du2018inversecsg, sharma2018csgnet}. However, approximating complex blends and free-form surfaces increases the number of primitives and parameters, and the recovered tree is not guaranteed to correspond to the feature structure intended by a designer. B-rep reconstruction can accurately represent boundary geometry, but surface reconstruction alone does not recover feature histories or dimensional dependencies. Learning-based CAD command reconstruction produces generative sequences, but it depends on large-scale training data and provides limited assurance that the reconstructed parameters correspond to dimensions required for practical design changes. Recent LLM-based approaches can generate CAD geometry or code from images or textual representations \cite{li2025llm4cad, badagabettu2024query2cad}. However, when design-intent discrimination and numerical or coordinate generation are combined within a single probabilistic process, it becomes difficult to maintain consistent parameter provenance and dependencies.
The common limitation of these approaches is not simply the absence or presence of parameters, but that they do not sufficiently address whether the reconstructed parameters can actually be used for design modification. Reverse parametrization should therefore aim not to assign arbitrary parameters to static geometry, but to explicitly construct a small set of dimensions corresponding to design intent, together with their dependencies and valid editing ranges, while preserving the default geometry. When the input is not one part but the set of parts that make up a product--given as separate, history-free solids with no record of how they fit together--the same requirement extends to the relations between them: the assembly relations must be recovered from the geometry alone, the dimensions that two parts share must be tied together, and the valid editing range must be the range within which the parts still assemble.

This study proposes a reverse-parametrization framework that combines LLM-based design-intent discrimination with a rule-based code emitter. Rather than simply reproducing the source geometry, the framework constructs an editable counterpart that preserves the original shape at its default parameter settings while allowing selected dimensions to vary according to the inferred design intent. The framework explicitly separates tasks requiring contextual judgment from deterministic geometric operations. A geometric kernel directly parses the input B-rep and serializes its geometric, topological, and relational information into a multichannel representation that can be interpreted by an LLM. Based on this representation, the LLM determines which dimensions should be exposed as independently editable design variables, which should remain fixed, and which should be defined as dependent on other dimensions, while also providing a rationale for each decision. A rule-based emitter then maps the discrimination results and dependency relations to predefined Grasshopper functions and constructs an executable parametric definition. Because all geometric values and modeling operations are obtained from the geometric kernel and predefined rules rather than generated directly by the LLM, this separation prevents unverified, LLM-generated geometric values from being directly introduced into the resulting model. In a subsequent validation stage within the generation-and-validation loop, the candidate parametric definition is evaluated to determine whether it preserves the source geometry at its default settings and remains geometrically valid under parameter changes. When validation fails, the results are fed back to revise the candidate definition or its exposed editing space. The same separation of roles is then applied between parts: the kernel measures where parts can be joined, the LLM confirms which of those relations the product actually has, and the kernel alone turns the confirmed relations into placements, couplings between dimensions, and verified editing ranges within one canvas. Design-intent reflection is evaluated separately through the structure and consistency of parameter-role assignments, the realization of selected dimensions as editable variables, and responsiveness to user-specified editing requirements. The framework requires neither a task-specific training dataset nor additional model training or fine-tuning. This training-free formulation is motivated by the practical difficulty of constructing large-scale geometry–intent pairs, as the design intent of each part must be established through expert judgment.

The proposed framework is demonstrated through the conversion of history-free B-rep models of mechanical parts, provided in STEP format, into editable Grasshopper definitions. In Grasshopper, dimensions selected for editing are exposed as Number Sliders, while their dependency relations are encoded through directed wiring and functional components. This representation makes the editability considered in this study directly observable and assessable.

A clear separation between contextual judgment and deterministic execution governs the framework architecture. The LLM performs the reasoning required to interpret design intent and determine whether each candidate dimension should be independently editable, remain fixed, or be defined as dependent on another dimension. In contrast, geometric measurement, serialization, and code emission are handled by the geometric kernel and rule-based emitter. Once the parameter roles and their dependency relations have been determined, constructing the corresponding Grasshopper definition becomes a predefined mapping task rather than an open-ended code-generation problem. The LLM is therefore employed only where contextual interpretation is required, while the construction of executable geometry remains under deterministic control. At the product level the LLM's contribution is narrower still: it selects among joint candidates the kernel has already measured, and everything that follows from that selection is computed.

This capability provides a practical bridge between static, history-free geometry and iterative engineering design. The framework may also be applicable to history-free B-reps obtained from topology-optimized geometry, subject to the supported reconstruction operations. More broadly, geometry originating from industrial styling, 3D scanning, or 2D-to-3D reconstruction may also constitute a potential input, provided that it has first been converted into a valid B-rep. The main contributions of this study are summarized as follows:

\begin{enumerate}
\item In this study, an editable counterpart in reverse parametrization is operationally defined by three jointly necessary conditions: preservation of the default geometry, reflection of design intent, and geometric validity under parameter variation. Each condition is stated for a part and for the assembled product: the product must preserve the part geometry and satisfy its assembly relations at the default pose, design intent includes which dimensions are shared between parts and which parts are joined, and validity includes the absence of interference and the integrity of every joint. Design-intent reflection is assessed through the selection of independently editable and dependent dimensions, consistency across repeated runs, the realization of selected dimensions as editing controls, and responsiveness to user requests. This formulation evaluates whether the recovered parameters support meaningful design modification, rather than merely determining whether parameters are present.

\item We develop a multichannel serialization scheme that represents STEP B-rep geometry using complementary numerical, relational, topological, and visual information. The representation combines kernel-derived geometric feature measurements, geometric relations such as coaxiality and repetition, a compressed attributed adjacency graph, and multiview renderings. This organization preserves design-relevant structural cues while grounding geometric values in kernel-derived measurements. The contribution of each channel is examined through ablation experiments.

\item We implement a STEP-to-Grasshopper framework that combines LLM-based design-intent interpretation with deterministic geometric measurement, parameter binding, definition emission, and validation, without requiring task-specific training or fine-tuning. The resulting Grasshopper definition exposes selected design variables as Number Sliders and represents their dependencies through component logic and data-flow connections, providing a directly manipulable, software-native editing interface. The generated definitions are evaluated in terms of global and region-specific geometric fidelity, design-intent reflection through parameter selection, variable realization, and request responsiveness. Parameter sweeps and boundary refinement adjust slider ranges, and shared parameters and measured dependencies coordinate geometric changes across assembled components.

\item We extend the framework from parts to the product they form by a kernel-driven assembly stage in which the LLM only confirms assembly relations among kernel-measured candidates. From the confirmed relations the kernel infers the closed-loop and attached poses of all parts, measures how the width of each receiving slot must follow the thickness of the seated part (measured geometric sensitivities), emits shared sliders, measured links and live joint references into a single canvas, and sweeps every user-facing slider to the last value at which the product remains interference-free, kinematically closable, with its bolted assembly relations preserved.

\end{enumerate}

The input to the transformation considered in this study is a STEP B-rep solid with its feature history removed, and the outputs are the corresponding Grasshopper definition $G$ and the independently exposed design-variable set $P$. When all sliders are placed at their default values measured from the source model, the geometry generated by $G$ must faithfully reproduce the source geometry. Dimensions that should be independently editable according to design intent must be exposed as sliders, dimensions that should change together must be made dependent on their reference variables, and dimensions that should remain fixed must be locked to their measured values. Finally, the generated geometry must remain geometrically valid when the exposed variables are changed within their verified ranges, which for a product means that the parts still assemble. This study defines these requirements as default geometry preservation, intent reflection, and edit validity, respectively, and uses their joint satisfaction as the operational definition of an editable counterpart. If any of these conditions is not satisfied, the resulting model cannot be reliably used for iterative design modification. Section 3 formally defines these conditions and presents the framework used to satisfy them.

The remainder of this paper is organized as follows. Section 2 reviews related work. Section 3 describes the four stages of the proposed framework: serialization, design logic construction, verification, and product assembly. Section 4 presents experimental results for default-shape preservation and design-intent reflection, together with assembly-level editing examples and ablation experiments. Finally, Section 5 summarizes the conclusions, limitations, and directions for future research.

\section{Related Work}
This section reviews research on shape reconstruction and generation relevant to the reverse parametrization of non-parametric models, organized into five categories: inverse CSG, B-rep reconstruction, CAD command sequence inference, LLM-based CAD generation and editing, and assembly-level joint inference. These directions provide context for the geometric representations, feature structures, and design-intent interpretation used in our serialization and reconstruction framework. This categorization follows the primary output representation and inference approach, while allowing overlaps such as LLM-based command sequence inference. We discuss the contributions and scope of each category in relation to our work, focusing on the requirement for parametric models established in Section~1: providing dimensions and dependency relations that designers can use for practical editing.

\subsection{Inverse CSG}
Inverse constructive solid geometry (CSG) reconstructs the constituent elements and operation structure of an input shape by decomposing it into Boolean combinations of analytic primitives. InverseCSG combines primitive extraction with program synthesis to search for a CSG tree corresponding to the input shape \citep{du2018inversecsg}. Among learning-based approaches, CSGNet introduces a sequential model that converts shapes into CSG programs, while UCSG-Net learns layers of Boolean operations without supervision \citep{sharma2018csgnet,kania2020ucsg}. CAPRI-Net and D$^2$CSG address both reconstruction fidelity and representation compactness by learning primitive combinations, whereas ExtrudeNet uses sketch--extrude pairs to obtain decompositions closer to CAD modeling operations \citep{yu2022capri,yu2023d,ren2022extrudenet}. More recently, DiffCSG introduces differentiable rendering of a given CSG model, enabling the optimization of continuous parameters against target images and extending its use to image-based editing \citep{yuan2024diffcsg}.

These approaches describe shapes through constituent elements and operations, providing a basis for adjusting the numerical values associated with individual elements. However, when primitive positions and sizes are reconstructed independently, finer decompositions may increase the number of values that must be adjusted, requiring an additional step to relate them to dimensions used by designers. For example, when a bolt circle is reconstructed as individual cylinders, their positions and sizes must be linked so that a designer can control the hole count, pitch circle diameter (PCD), and hole diameter. 
Moreover, because multiple CSG structures can represent the same shape, even applying a decomposition suitable for geometric reconstruction does not guarantee that the mechanical assembly and geometric requirements are fully satisfied.  
Therefore, obtaining a parametric model under the definition assumed in this study requires constructing the design dimensions and their dependency relations, as well as performing primitive decomposition and numerical reconstruction.

\subsection{B-rep Reconstruction}
B-rep reconstruction estimates surfaces, boundaries, and topological connections from unstructured geometric data, such as point clouds and meshes, to obtain boundary representations usable by CAD kernels. Traditional reverse engineering approaches address tasks such as region segmentation, surface-type classification, and surface fitting \citep{benkHo2001algorithms}, while learning-based methods have increasingly integrated the estimation of surfaces and topology. ParSeNet reconstructs surface patches from point clouds, ComplexGen jointly predicts vertices, edges, faces, and their adjacency relations, and Point2CAD combines surface reconstruction with intersection-curve computation to reconstruct B-reps \citep{sharma2020parsenet,guo2022complexgen,liu2024point2cad}. For generation, SolidGen and BrepGen introduce approaches that directly generate B-rep geometry and topology \citep{jayaraman2022solidgen,xu2024brepgen}. BrepDiff represents faces as masked UV grids within a single-stage diffusion model, constructs solid B-reps through post-processing, and supports shape completion, merging, and interpolation \citep{lee2025brepdiff}.

These methods provide an important foundation for reverse engineering by reconstructing boundary geometry and generating geometric and topological structures that can be processed by CAD kernels. Although the resulting boundary expressions can be edited directly and the geometry can be modified within the authoring system, this form of editability should not be confused with the recovery of a parametric design representation. In particular, existing approaches generally do not explicitly distinguish meaningful engineering design variables from numerical parameters introduced solely to approximate the geometry. This limitation is especially evident when dense meshes reconstructed from point clouds are converted into B-rep models. Although such models may represent the target shape with high geometric accuracy, their large number of low-level geometric entities does not provide an intuitive or compact means of modifying the shape. In this respect, they remain similar to non-parametric models.
For methods primarily concerned with boundary reconstruction, recovering such driving dimensions and their dependencies requires an additional parametrization process. The present work takes these boundary representations as its starting point and constructs an editable variable structure that reflects meaningful design parameters and their relationships.

\subsection{CAD Command Sequence Inference}
CAD command sequence approaches represent shapes through ordered modeling commands and operation-specific parameters, either reconstructing these sequences from input geometry or generating new ones. Fusion 360 Gallery provides human-authored modeling sequences and an environment for their execution, while ABC provides a foundation for geometric learning through curve, surface, and boundary geometry data \citep{willis2021fusion,koch2019abc}. DeepCAD introduces a Transformer-based approach for generating sequences of sketch and extrusion commands, and SkexGen improves control over generation through separate codebooks for topology, geometry, and extrusion \citep{wu2021deepcad,xu2022skexgen}. Approaches that recover modeling operations from geometry include Point2Cyl for point-cloud-based extrusion decomposition, CAD-SIGNet for sketch--extrude sequence inference, and CAD-Recode for executable CadQuery code generation \citep{uy2022point2cyl,khan2024cad,rukhovich2025cad}. Using B-reps directly as input, eCAD-Net addresses the reconstruction of editable CAD models, while CADCL reconstructs command sequences through contrastive learning between B-rep and CAD sequence representations \citep{zhang2025ecad,liang2025cadcl}.

Recent research has explored not only broader command vocabularies but also explicit representations of constraints and references among modeling commands. HistCAD introduces a history-based representation and dataset that include sketch constraints, feature operations, and boundary references, and evaluates editing success and constraint preservation after parameter changes \citep{dong2025histcad}. Constraint-solver feedback has also been used to improve sketch constraint generation, increasing the proportion of fully constrained sketches while reducing geometric distortion \cite{casey2025aligning}. These studies indicate a progression from reconstructing executable command sequences toward representations that incorporate constraints and reference relationships during editing. However, the recovery of an executable command sequence does not necessarily determine which dimensions should serve as independent driving variables for a particular design task. The present study instead focuses on constructing an alternative parametric definition from a history-free B-rep, in which selected dimensions are assigned independent, fixed, or dependent roles according to design intent.

\subsection{LLM-Based CAD Generation and Editing}
The use of LLMs and VLMs for CAD generation and editing has grown rapidly. Following early studies on generating CAD code from natural language, Text2CAD addresses text-conditioned command sequence generation, while CAD-MLLM explores multimodal CAD generation using text, images, and point clouds \citep{makatura2024can,khan2024text2cad,xu2024cad}. LLM4CAD evaluates CAD generation using textual and visual information, and Query2CAD introduces iterative refinement based on visual feedback from generated results \citep{li2025llm4cad,badagabettu2024query2cad}. CAD-Assistant extends these approaches through an agent architecture that uses CAD-related tools and APIs \citep{mallis2025cad}. For editing, CAD-Editor identifies the portions to modify according to natural-language instructions and regenerates them, while B-repLer performs language-guided shape editing in a learned B-rep latent space without requiring a modeling history \citep{yuan2025cad,liu2025b}.

These studies expand CAD automation through natural-language-based generation and editing, tool use, and feedback-driven refinement. Within this broader direction, our work focuses on constructing a variable structure that designers can repeatedly use to edit existing STEP geometry without a modeling history. To this end, we serialize geometric and relational information from the B-rep, determine whether dimensions should be independent, fixed, or dependent, and use a rule-based emitter to construct the corresponding variables and connections in Grasshopper. We also evaluate both default-shape preservation and geometric validity after parameter changes, using verification results to guide subsequent revisions.

\subsection{Assembly-Level Joint Inference}
Research on mechanical assemblies has studied how individual parts connect and are constrained. Learning-based joint-inference methods predict joint axes, joining surfaces, or joint types between part pairs from their B-rep geometry, typically using assembly datasets with annotated joint relationships \cite{jones2021automate, willis2022joinable}. Assembly-generation methods also construct multiple parts and their assembly relations. For example, AssemCAD converts natural-language descriptions into explicit part, interface, and mating specifications and realizes them through deterministic geometric operations \cite{dong2026assemcad}. Constraint-based approaches represent relationships between assembly features, solve spatial constraints between parts, and identify the remaining kinematic degrees of freedom \cite{anantha1996assembly}.

These studies show that joints can be inferred from part geometry and used to determine component relative placement. For reverse parametrization of history-free parts, an additional requirement is to determine how dimensions in connected components should change together while preserving assembly relationships. For example, changing the thickness or length of one component may require corresponding adjustments to a receiving slot, bolt-hole locations, or other interfacing features. It may also make the assembly geometrically infeasible or prevent the mechanism from reaching a valid configuration.
The present work addresses these relationships from the perspective of assembly-level parametrization. We first identify joint candidates from the reconstructed part geometries using the geometric kernel. The LLM then determines which candidate relationships correspond to the intended assembly structure. The kernel then translates each confirmed relationship into an explicit, editable representation, such as a shared design variable, a dimension-dependency relation, or a parameterized relative placement, along with a valid operating range. Thus, joint inference is not the final output but the basis for constructing an assembly-level parametric representation.

\section{Methodology} \label{sec:method}

\begin{figure}
\centering
 \includegraphics[width=1\textwidth]{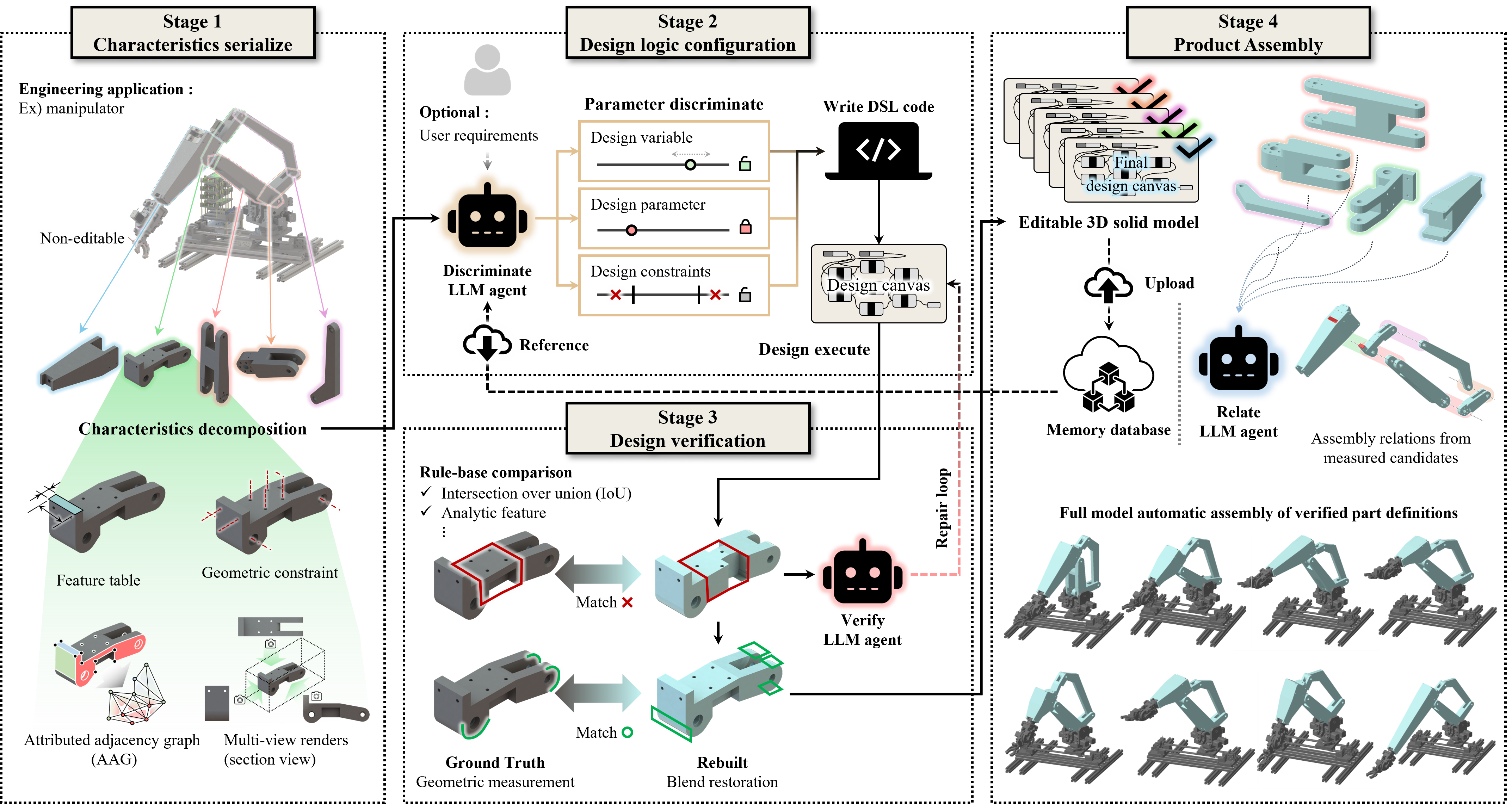}
\caption{Framework of the proposed method.}
\label{fig:framework}
\vspace{-9pt}
\end{figure}

This study proposes a framework to convert a set of STEP B-rep parts without feature histories or assembly information into a single Grasshopper definition that can be edited via key dimensions aligned with design intent. The framework consists of four stages: geometry serialization, design logic construction, design verification, and product assembly (Figure~\ref{fig:framework}). The first three stages operate on each part, and design logic construction comprises two subprocesses: LLM-based discrimination and rule-based emission. The fourth operates between parts and applies the same three steps to the relations among them. The fundamental principle of this framework is role separation. The LLM handles tasks requiring open-ended judgment, such as interpreting design intent and confirming which measured contacts are assembly relations. In contrast, the geometric kernel and rule-based modules handle tasks requiring reproducible results, including geometric measurement, feature graph construction, pose inference, Grasshopper canvas emission, and quantitative verification. No additional model training is involved.

\subsection{Problem Formulation and Editability Criteria}\label{sec:problem}

First, specify the target parametric representation for a STEP model without a feature history. This study adopts Grasshopper, the parametric modeling environment of Rhinoceros 3D\textsuperscript{\textregistered}. Editable dimensions appear as explicit Number Slider objects, and connections represent dependencies between dimensions. A domain-specific language with a small, predefined command set can automatically construct a canvas, and the resulting geometry can be exported as STEP. Canvas construction and verification against the source can therefore be performed within the same execution workflow. The input is a set of STEP solids $\{S_i\}_{i=1}^{m}\subset\mathcal{B}$ without feature histories and without any assembly information--no joints, no constraints, no relative placement--and the output consists of a Grasshopper definition $G$ with connected components and Number Sliders, together with the set of exposed design variables $P$. For a single part, we regard $G$ as a partial function whose value is defined when execution within the constrained design variable space succeeds and produces a single B-rep solid (in Eq.~\ref{eq:gh-map}); for a product, $G$ returns one placed solid per part, $T_i(\theta)\,G_i(\theta)$, where $G_i$ is the part-level definition of part $i$ and the placement $T_i$ is itself a function of the sliders.


\begin{equation}
G:\Theta_C\rightharpoonup\mathcal{B},\qquad \theta\mapsto G(\theta).
\label{eq:gh-map}
\end{equation}

Here, $\theta$ is a vector containing the values of $n$ independent sliders--the part sliders that survive discrimination, the sliders shared between parts, and, for a mechanism, its pose inputs--and $\Theta=\prod_{i=1}^{n}[l_i,u_i]$ is the design space defined by their declared ranges. $\Theta_C$ denotes the subspace satisfying the relational constraints $C$, and $\theta_0$ is the vector in which all sliders are set to their default values measured from the STEP model. An editable counterpart is defined by the following three conditions.

\paragraph{(i) Default-shape preservation.}
When all sliders are set to their default values, the shape $\widehat S=G(\theta_0)$ generated by $G$ must faithfully reproduce the source STEP solid $S$. Because global volumetric overlap alone may not adequately detect missing small analytic features, we use a combined criterion incorporating global fidelity and region-specific metrics (in Eq.~\ref{eq:default-match}).

\begin{equation}
\begin{gathered}
\mathrm{MATCH}_{0}(\widehat S,S)
=\mathbb{I}[\operatorname{IoU}_{3D}\ge\tau_g]
\land\mathbb{I}[\operatorname{IoU}_{\mathrm{af}}\ge\tau_{\mathrm{af}}]\\
\land\mathbb{I}[r_{\mathrm{org}}\le\tau_{\mathrm{org}}]
\land\mathbb{I}[e_{\mathrm{breach}}\le\tau_e]\\
\land\mathbb{I}[n_{\mathrm{missing}}=0].
\end{gathered}
\label{eq:default-match}
\end{equation}

$\operatorname{IoU}_{3D}$ measures overall shape fidelity, $\operatorname{IoU}_{\mathrm{af}}$ measures fidelity in analytic-feature regions, $r_{\mathrm{org}}$ is the rebuilt-to-source occupied-volume ratio in irregular body regions, $e_{\mathrm{breach}}$ is the boundary-breach ratio in analytic-feature regions, and $n_{\mathrm{missing}}$ is the number of missing hole features. For a product, condition (i) is required of every part and of the assembly: the rebuilt parts must satisfy part-local shape matching and the interpenetration limit at the inferred poses (in Eq.~\ref{eq:default-match-product}).

\begin{equation}
\mathrm{MATCH}_{0}^{\mathrm{prod}}
=\bigwedge_{i}\mathrm{MATCH}_{0}(\widehat S_i,S_i)\;\land\;
\bigwedge_{i<j}\mathbb{I}\bigl[V\bigl(T_i\widehat S_i\cap T_j\widehat S_j\bigr)\le\tau_{\cap}\bigr].
\label{eq:default-match-product}
\end{equation}

\paragraph{(ii) Design-intent reflection.}
Although an effectively unlimited number of dimensions can be extracted from a B-rep solid, only a small subset is meaningful for design. This condition requires those dimensions to be incorporated into $G$ with the appropriate roles. Dimensions that should be adjusted independently according to design intent must be exposed as independent sliders in $P$, dimensions that should vary together must depend on a reference variable, and the remaining dimensions must be fixed at their measured values. Identifying these roles among numerous candidates is the task of the discrimination stage (Section~3.3). For a product, the condition has two further parts: dimensions that must vary together across two parts--such as a pin diameter and the counterpart bore diameter, or a slot width and the thickness of its inserted tab--must be represented by a shared slider or a measured link that preserves the required clearance, and the set of assembly relations that holds the product together must be the one a designer would recognize. Identifying the shared dimensions is again discrimination; identifying the joints is the task of the joint-confirmation step of Stage~4 (Section~\ref{sec:assembly}).

Without this condition, geometric accuracy alone does not establish that the reconstructed parameters support the intended design changes. Section~\ref{sec:experiments} evaluates this condition through the structure and run-to-run consistency of discrimination results, the promotion process through which selected dimensions become implemented variables, and responsiveness to designer requests specifying particular dimensions.

\paragraph{(iii) Edit validity.}
The geometry generated within the exposed editing space must remain geometrically valid. Even if the declared range of each slider is individually valid, simultaneous changes to multiple variables may eliminate the wall between holes or cause two hub profiles to overlap. We therefore aim for $G(\theta)$ to be valid for every $\theta$ in the design space $\Theta_C$ satisfying the relational constraints $C$. The validity of the geometry generated at a particular slider configuration $\theta$ is defined as the conjunction of five deterministic predicates.

\begin{equation}
\begin{gathered}
\mathrm{valid}(G(\theta))
=\mathrm{built}
\land(\#\mathrm{bodies}=1)
\land\mathrm{isSolid}
\land\mathrm{isValid}\\
\land\bigl[V(G(\theta))>\varepsilon_V V_0\bigr].
\end{gathered}
\label{eq:validity}
\end{equation}

$\mathrm{built}$ indicates successful Grasshopper solution and output generation, or equivalently that the part-level function is defined at the tested input. The predicate $\#\mathrm{bodies}=1$ checks whether the result remains a single body. $\mathrm{isSolid}$ indicates whether the result is a closed solid, while $\mathrm{isValid}$ denotes the internal geometric and topological validity of the B-rep as determined by the geometric kernel. The volume lower bound excludes degenerate results that are formally valid but have effectively disappeared, where $V_0=V(G(\theta_0))$ is the volume of the default shape. At the product level, the target condition is $\bigl(\bigwedge_i\mathrm{valid}(G_i(\theta))\bigr)\land\mathrm{assembled}(G(\theta))$. The implemented sweep checks successful rebuilding of each part and the assembly condition in Eq.~\ref{eq:validity-product}. The remaining part-level predicates are evaluated during part-level verification and are not reevaluated by the product sweep; passing the product test therefore does not independently establish them at every edited product configuration. Assembly conditions require closure of any kinematic loop, interpenetration depth below $\tau_d$ and preservation of the confirmed bolted joints. Pin-joint positions are maintained by the loop solution. The implemented product-level test is therefore $\bigl(\bigwedge_i\mathrm{built}_i(\theta)\bigr)\land\mathrm{assembled}(G(\theta))$.

\begin{equation}
\mathrm{assembled}(G(\theta))
=\mathrm{closable}(\theta)\;\land\;
\bigwedge_{i<j}\bigl[d_{\cap}\bigl(\widetilde S_i(\theta),\widetilde S_j(\theta)\bigr)\le\tau_d\bigr]\;\land\;
\bigwedge_{(a,b)\in\mathcal{J}_{\mathrm{bolt}}}\mathrm{joined}_{ab}(\theta).
\label{eq:validity-product}
\end{equation}

Here, $\widetilde S_i(\theta)=T_i(\theta)G_i(\theta)$ denotes the rebuilt geometry of part $i$ placed at its current assembly pose. To handle this condition through four part-level procedures and two more at the product level. Discrimination declares relations between dimensions that should vary together; the kernel derives conditions preventing geometric collision or separation from the feature graph and measurements; the kernel computes the numerical limits of these conditions and injects them into the Grasshopper canvas; and single-axis sweeps and multivariable exploration on the completed canvas apply Eq.~\eqref{eq:validity} to the actual editing results. At the product level, the kernel converts each measured seated contact into a dimension link, so an edit to one part is followed by the slot that receives it. It sweeps every user-facing slider on the merged canvas, applies Eq.~\ref{eq:validity-product}, and trims the slider to the last value at which the product still assembles (Section~\ref{sec:assembly-sweep}). However, exhaustively inspecting a continuous, high-dimensional editing space is infeasible. This study therefore limits its empirical scope to the intervals assessed by single-axis sweeps and the multivariable configurations sampled through corner and interior exploration.

Each of the three conditions is independently necessary. Omitting any one yields a model unsuitable for editing; satisfying all three therefore provides an operational definition of the editability criterion established in Section~1. Condition~(i) is addressed by deterministic emission, shape-fidelity verification, and, for a product, kernel pose inference with interference checking; condition~(ii) by variable discrimination and joint confirmation; and condition~(iii) by the computation and injection of validity constraints, measured seat links, and empirical verification of the editing space at both levels.

\subsection{Stage 1: Geometry Serialization --- A Multichannel B-rep Representation}
\label{sec:serialization}

Serialization $\Sigma(S)$ follows two principles. First, it maintains a single source of precise numerical information. Dimensions and coordinates used for geometry generation come from kernel measurements in Channel~1, while the other channels provide structural, relational, and visual context. Auxiliary numerical values in those channels, such as face areas, also originate from the same kernel extraction results. Second, arithmetic that derives new values from declared values is not delegated to the LLM. The kernel precomputes distances that can be obtained through finite combinations, while additional dimensions not included in the lists are acquired through the measurement protocol in Section~\ref{Stage_2}.

\subsubsection{Channel 1 --- feature table} \label{Channel_1}

Channel 1 consists of two layers. The first layer, the feature table, lists analytic surfaces collected by traversing the B-rep faces using OpenCASCADE. For cylindrical faces, the radius, axis direction, axis location, and axial span are measured, and holes and bosses are classified using the relationship between the face normal $\mathbf n$ and the radial direction $\mathbf r$. Coaxial cylindrical faces with equal radii are merged into a single feature. For conical faces, the angle and reference radius are recorded, while geometrically equivalent planar faces are consolidated. For prismatic parts with a constant extrusion cross-section, a polyline of profile vertices obtained from the cross-section is also recorded to provide coordinates for the body silhouette. Each feature and face receives an identifier that can be referenced consistently within a run, and the measurement protocol resolves these identifiers to the actual B-rep entities.

Kernel-computed derived distances are appended to the table in two blocks. First, the exhaustive distance table is an uninterpreted lookup table containing all possible pairs within the categories defined for querying in this study: separations between planes along the same axis, distances between feature centers, and distances between parallel axis lines. Second, the rule-based length list is not a reduced version of the distance table. Instead, it provides a separate interpretation layer that selects gaps, wall thicknesses, and inter-axis distances to which structural meaning can be assigned. When the normals of two parallel planes face each other, the intervening void is classified as a candidate gap; when they point away from each other, the intervening material is classified as a candidate wall thickness. The perpendicular distance between parallel axis lines is recorded as an inter-axis distance. For gaps and wall thicknesses, exterior-boundary pairs are excluded and representative entries are retained. Inter-axis distances are recorded for the detected pairs of parallel axis lines. Although these rules are deterministically reproducible, they do not fully determine design meaning and may produce extra or missing candidates. The list therefore ends with instructions to obtain other lengths through the measurement protocol. All numerical values in both blocks come from geometric kernel measurements or computations; the LLM calculates none.

The second layer, the feature graph, represents the same geometry at the feature level for reconstruction. While the feature table lists the constituent faces and their measurements, the feature graph specifies how the measured geometry is reconstructed using the limited set of build modules in this study. Bosses and bores are recorded through their center coordinates and axial spans, while bolt holes repeated around a circle are compressed into a single pattern defined by count, pattern radius, and phase. A top-view projection polyline represents the body silhouette. Height intervals separated by planes perpendicular to the principal hole axis are defined as tiers. The height range of each tier and the features passing through it provide the numerical basis for extrusion and cutting depths during emission. Although this layer also supports discrimination, its primary consumer is the emission stage. For discrimination, the feature graph grammar directly presents dimensions as design variable candidates. Dimensions outside the grammar cannot immediately be declared as design variables because their candidate identifiers and connections to build modules have not been defined.

\subsubsection{Channel 2 --- geometric constraints}\label{Channel_2}

Channel 2 declares the structure among dimensions. A region tree partitions the part into intervals along its dominant axis and assigns each feature to its region. Coaxial groups indicate that a stepped hole forms a single stack rather than several independent cylinders. Mirror symmetry and parallel axis pairs identify candidate hinge or pivot structures, followed by the detection of bolt-circle and counterbore combinations. In particular, a bolt circle summarizes the coordinates and diameters of $n$ holes as three dimensions: count, PCD, and hole diameter. These dimensions directly become slider candidates, reversing the tendency of primitive-based CSG representations to increase parameters with the number of primitives rather than with the design degrees of freedom. This channel contains only identifiers and structure; all values are stored in Channel~1. Its geometric constraints declare structural relations present in the current geometry---which features share an axis, form a pattern, or exhibit symmetry---and provide structural evidence for subsequent LLM-based discrimination (in Section~\ref{Stage_2}).

\subsubsection{Channel 3 --- AAG topology tree}\label{Channel_3}

Channel 3 constructs an attributed adjacency graph (AAG) with B-rep faces as nodes and shared edges between faces as graph edges \cite{joshi1988graph}. Each node stores its surface type and area, and each graph edge is assigned one of the labels \textsf{smooth, convex, concave} to describe the relationship between its adjacent faces. The labels are determined deterministically from the two outward normals and the edge tangent along the shared edge. The graph is compressed into a textual topology tree through four transformations. First, coaxial cylindrical faces with matching radius, axis direction, axis location, and classification are merged into a single feature node; Figure~\ref{fig_aag} shows this step for the rocker, where the 54 faces and 130 adjacencies of the raw AAG reduce to 39 nodes and 93 adjacencies once the shaded coaxial groups are merged. Second,  nonplanar B-spline faces connected by \textsf{smooth} edges into a single tangential patch. Third, remaining nodes are removed if their areas fall below the larger of an absolute area threshold and a relative threshold based on total area. Analytic-feature nodes composed of cylindrical or conical faces and the body root node are protected regardless of area. Fourth, the compressed graph is converted into a tree using breadth-first search (BFS) for textual serialization. The node with the largest area becomes the root. The resulting hierarchy follows graph distance from the root, providing a compact textual representation. Nodes at the same depth are not necessarily siblings or symmetric features; patterns and symmetry are therefore interpreted together with the separate geometric relation channel.


\begin{center}
\includegraphics[width=0.8\textwidth]{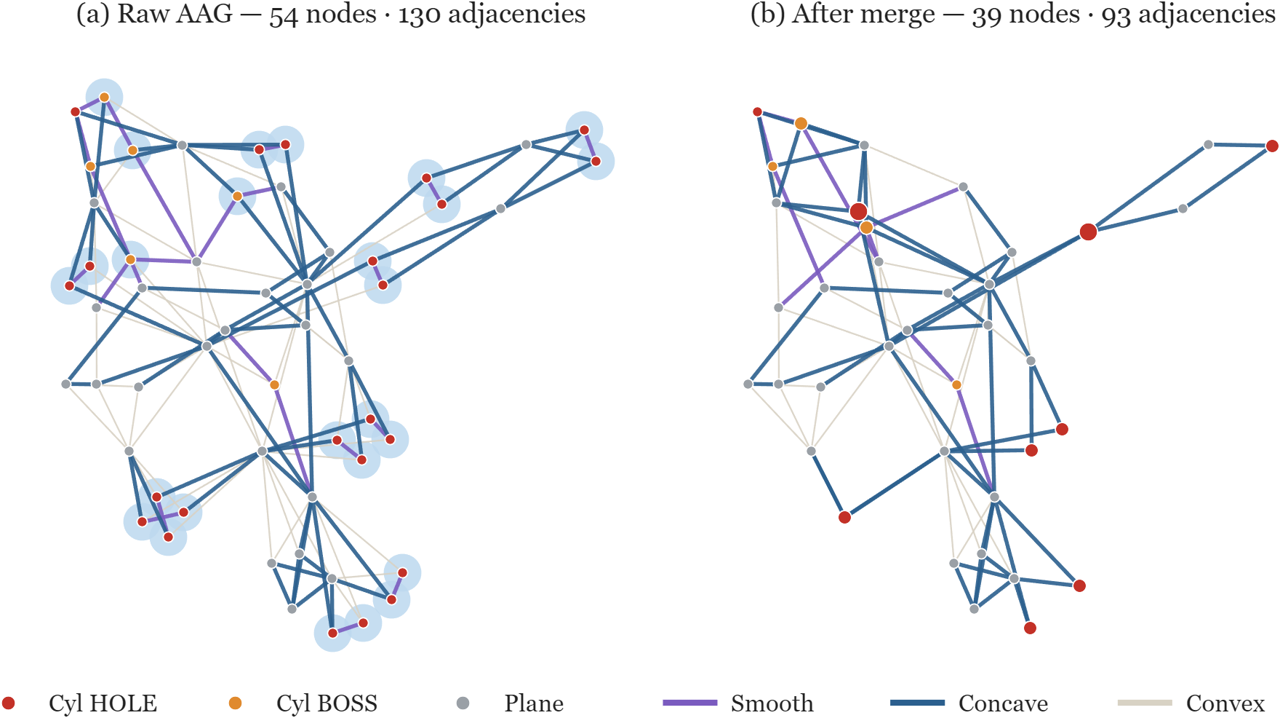}
\captionof{figure}{Attributed adjacency graph of the rocker. (a)~Raw AAG: 54 face nodes and 130 adjacencies; shaded discs mark the coaxial groups that are merged. (b)~After coaxial merging: 39 nodes and 93 adjacencies. Node color encodes surface type; edge color encodes the dihedral label of the shared edge.}
\label{fig_aag}
\end{center}

\subsubsection{Channel 4 --- multiview renderings}\label{Channel_4}

The framework generates shaded multiview renderings using the view names and orientation conventions of the Rhino viewport in which the Grasshopper geometry is displayed. It also generates cross-sectional views perpendicular to the principal hole axis, 2D profile drawings for prismatic parts, and renderings color-coded according to the semantic labels derived from the AAG. Each 3D view includes a coordinate triad indicating its orientation. These images provide visual context for verifying orientation and consistency across the serialized channels. All numerical coordinates used in subsequent stages are obtained exclusively from geometric kernel measurements rather than from the images.

\subsubsection{Serialization example}\label{Serialization_example}
\label{sec:serial-example}

Figure~\ref{fig:rocker} and Tables~\ref{tab:features}--\ref{tab:relations} illustrate the outputs of Channels~1 and~2 for the rocker, which serves as a running example throughout this paper. Channel~1 lists 13 cylindrical features together with their kernel-measured diameters, axial lengths, axis directions, and axis locations (Table~\ref{tab:features}). Based on this measured geometry, Channel~2 identifies two coaxial groups, one stepped bore, one pair of parallel axes, one rectangular hole pattern, and the inter-axis distance between the two bores (Table~\ref{tab:relations}). These serialized records are provided to the LLM as text. The feature labels overlaid on Figure~\ref{fig:rocker} are included solely to aid reader interpretation and are not part of the LLM input.

\begin{center}
\includegraphics[width=0.5\textwidth]{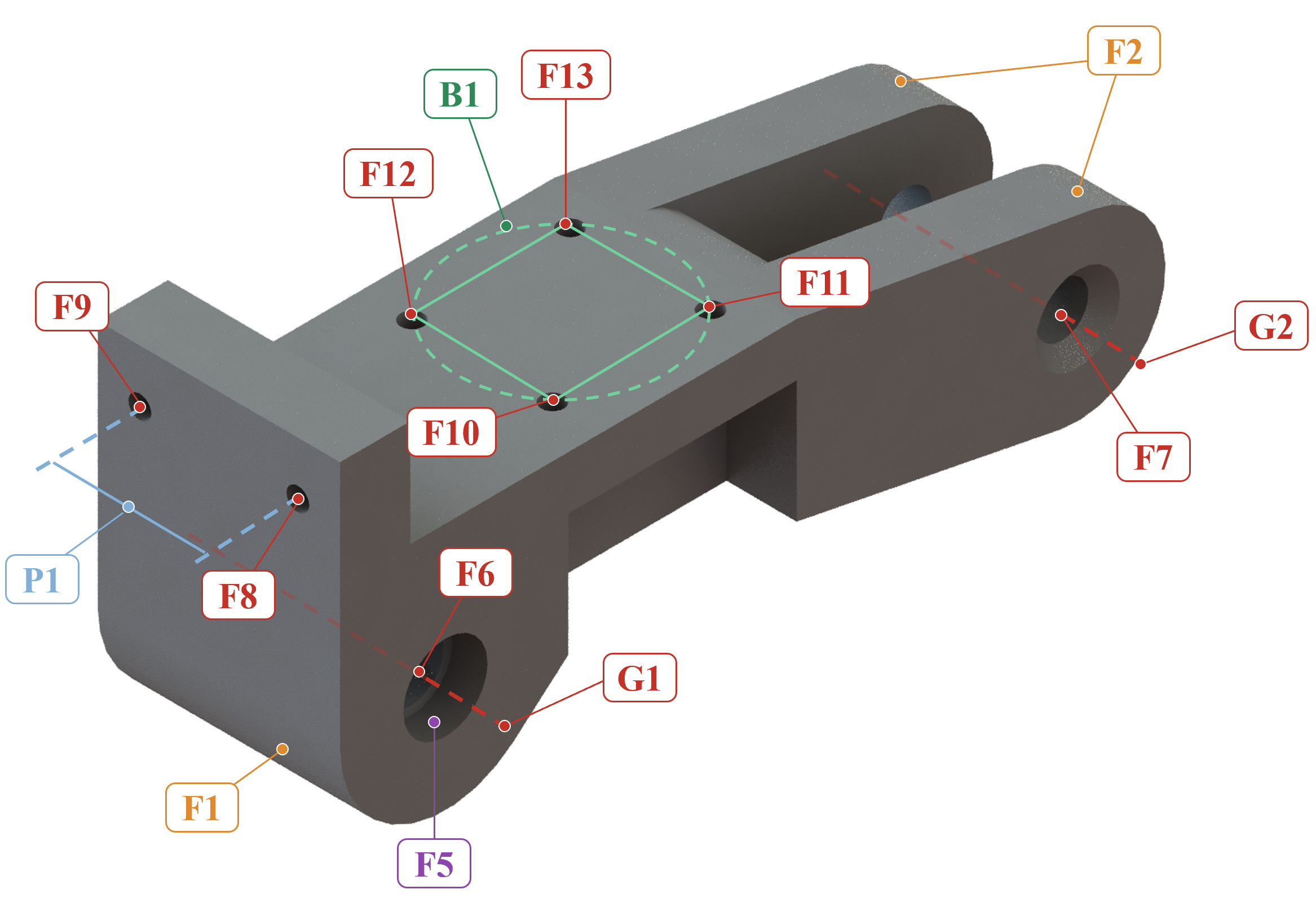}
\captionof{figure}{Running example \texttt{Rocker\_Body}. Cylindrical features \fid{F1}--\fid{F13} from Channel~1 (red: hole, orange: boss, purple: counterbore) and the relations from Channel~2 recovered from their axes: coaxial groups G1 and G2 (dashed axes), parallel pair P1, and hole pattern B1. Feature attributes are listed in Table~\ref{tab:features} and relations in Table~\ref{tab:relations}.}
\label{fig:rocker}
\end{center}

\begin{table}[htbp]
\centering\small
\caption{Channel 1 output for \texttt{Rocker\_Body} (Fig.~\ref{fig:rocker}): cylindrical features, kernel-measured. $L$: axial length; coordinates in the min-corner frame, dash = along the axis.}
\label{tab:features}
\setlength{\tabcolsep}{5pt}
\begin{tabular}{@{}l l r r c l l@{}}
\toprule
ID & Type & $\dia$ & $L$ & Axis & Axis point & Grp \\
\midrule
\fid{F1}  & boss             & 60.0 & 69.0 & $Z$ & (30, 30, --)    & G1 \\
\fid{F3}  & boss (fork round) & 40.0 & 29.0 & $Z$ & (132, 49, --)   & -- \\
\fid{F4}  & boss (fork round) & 40.0 & 29.0 & $Z$ & (131, 40, --)   & -- \\
\fid{F6}  & hole, thru       & 19.0 & 59.0 & $Z$ & (30, 30, --)    & G1 \\
\fid{F5}  & c'bore $\times 2$ & 24.0 &  5.0 & $Z$ & (30, 30, --)    & G1 \\
\fid{F2}  & boss $\times 2$   & 50.0 & 20.0 & $Z$ & (210, 30, --)   & G2 \\
\fid{F7}  & hole $\times 2$   & 15.0 & 20.0 & $Z$ & (210, 30, --)   & G2 \\
\fid{F8}  & hole, blind      &  6.5 & 20.0 & $X$ & (--, 85, 57)    & P1 \\
\fid{F9}  & hole, blind      &  6.5 & 20.0 & $X$ & (--, 85, 12)    & P1 \\
\fid{F10} & hole, blind      &  6.5 & 15.0 & $Y$ & (75, --, 54.5)  & B1 \\
\fid{F11} & hole, blind      &  6.5 & 15.0 & $Y$ & (120, --, 54.5) & B1 \\
\fid{F12} & hole, blind      &  6.5 & 15.0 & $Y$ & (75, --, 14.5)  & B1 \\
\fid{F13} & hole, blind      &  6.5 & 15.0 & $Y$ & (120, --, 14.5) & B1 \\
\bottomrule
\end{tabular}
\end{table}

\begin{table}[htbp]
\centering\small
\caption{Channel 2 output for \texttt{Rocker\_Body}: relations derived from the feature axes in Table~\ref{tab:features}. Coaxiality: $|\hat u_i\!\cdot\!\hat u_j|>0.999$, $\varepsilon_R=0.05$, $\varepsilon_d=0.2$~mm.}
\label{tab:relations}
\setlength{\tabcolsep}{5pt}
\begin{tabular}{@{}l l l@{}}
\toprule
Relation & Members & Evidence \\
\midrule
Coaxial G1   & \fid{F5}, \fid{F6}, \fid{F1}        & one axis $\parallel Z$ through (30, 30) \\
Coaxial G2   & \fid{F7}, \fid{F2}                  & one axis $\parallel Z$ through (210, 30) \\
Stepped bore & \fid{F5}$\to$\fid{F6}$\to$\fid{F5}  & $\dia24$ (5) / $\dia19$ (59) / $\dia24$ (5) \\
Parallel P1  & \fid{F8}, \fid{F9}                  & axes $\parallel X$, pitch 45.0 \\
Pattern B1   & \fid{F10}--\fid{F13}                & $4\times\dia6.5$, $45\times40$~mm (PCD 60.2) \\
Axis--axis   & G1 $\leftrightarrow$ G2             & 180.0~mm \\
\bottomrule
\end{tabular}
\end{table}

\subsection{Stage 2 (a): Variable Discrimination}\label{Stage_2}
\label{sec:discrimination}

The discrimination problem is formulated as follows. Given the serialized multichannel representation $\Sigma(S)$ and the slider candidate set $P_{\mathrm{cand}}$, the LLM produces an editing-state assignment for each candidate, intent-based couplings, and natural-language rationales for its decisions.

\begin{equation}
\mathcal{D}\!\left(\Sigma(S),P_{\mathrm{cand}},\mathcal{R}\right)
\rightarrow
\left(\mathcal{V},C_{\mathrm{int}},\rho\right),
\label{eq:discrimination}
\end{equation}

Here, $\mathcal{D}$ is the discrimination agent, and $\mathcal{R}$ is a reference block of retrieved discrimination precedents. When precedent memory is not used, $\mathcal{R}=\varnothing$. $\mathcal{V}$ classifies each candidate as \textsf{OPENED}, \textsf{FIXED}, or \textsf{MIRROR}$(t)$; $C_{\mathrm{int}}$ represents intent-based couplings between variables that should vary together; and $\rho$ provides a natural-language rationale for each decision. Specifically, \textsf{OPENED} exposes a dimension as an independent slider, \textsf{FIXED} locks it at its measured value, and \textsf{MIRROR(t)} declares a dependency on the reference parameter $t$. Neither \textsf{FIXED} nor \textsf{MIRROR} adds an independent slider, but the latter connects the dependent dimension to the reference variable so it changes with it.

Validity constraints $C_{\mathrm{val}}$ describe geometric facts about the measured shape rather than design intent. They are therefore derived directly by the kernel rather than included in the output of Eq.~\eqref{eq:discrimination}. The constraint set $C$ introduced in Section~\ref{sec:problem} combines the two sources as $C=C_{\mathrm{int}}\cup C_{\mathrm{val}}$. Thus, the discrimination specification does not supply new numerical values for geometry construction. The LLM declares only the state of each dimension ($\mathcal{V}$) and the relations that should vary together ($C_{\mathrm{int}}$); all default values and numerical constraint limits are supplied by kernel measurements and computations (Sections~3.2 and~3.4). The natural-language rationale $\rho$ does not modify geometry but makes the discrimination decisions auditable by recording the functional reasoning behind each assignment.

However, $P_{\mathrm{cand}}$ is a finite candidate set automatically derived from the kernel feature graph, so design dimensions outside its grammar are not presented as candidates. To accommodate such dimensions, the discrimination stage includes a measurement protocol (Section~3.3) that provides a promotion path. Let $P_{\mathrm{ext}}^{\mathrm{acc}}$ denote the promoted candidates that pass the acceptance gate (Section~3.5). Before product-level sharing and locking, the slider set is then $P=\mathcal{V}^{-1}(\textsf{OPENED})\cup P_{\mathrm{ext}}^{\mathrm{acc}}$.

When several parts are converted together, discrimination is performed once for all of them in a single call rather than once per part, so that a dimension of one part can be declared dependent on a dimension of another. Such a declaration is a role assignment with a rationale and carries no value; whether and how the two dimensions are tied together is decided afterwards by measurement (Section~\ref{sec:assembly-links}).

The judgment required by discrimination---which dimensions should be editable according to design intent---is difficult to reduce to geometric rules. Recognizing that a fork gap is an interface for the component it joins, that a coaxial stack forms a pivot axis, or that a particular diameter corresponds to a standard fastener requires knowledge of engineering conventions and function beyond the geometry itself, making these judgments difficult for rule-based code. This motivates using an LLM for discrimination: the stage uses a general-purpose LLM's pretrained knowledge to interpret functional context and common engineering conventions that geometry alone cannot readily infer. Because the framework cannot guarantee the accuracy of this knowledge, it treats LLM outputs as proposed discrimination decisions subject to review and consistency checks rather than definitive assignments.

The framework does not accept these judgments without review. Instead, it provides three mechanisms to constrain their evidential basis and support human inspection. First, the kernel-generated four-channel representation grounds decisions in measured geometric information. Second, structurally similar previous decisions and user corrections are provided as reference precedents. Third, each decision includes a functional rationale $\rho$ so that a person can inspect its reasoning. These natural-language rationales support auditability, not evidence that automatically validates a decision.

The LLM-generated discrimination specification undergoes deterministic consistency checks before use. The checks verify that candidate identifiers and state labels belong to the allowed lists, that reference variables named by \textsf{MIRROR} exist, and that no candidate receives conflicting declarations. Conflicting or unresolved declarations are revised or rejected according to predefined arbitration rules, and both the original declaration and the reason for the change are recorded in an audit log. The permitted content of the specification is also restricted. To prevent unmeasured construction values from entering the model, the model prohibits arbitrary coordinates except for direct references to constants supplied in the channels.

This process corresponds to the retrieve--reuse--revise--retain cycle of case-based reasoning \cite{aamodt1994case}. First, the system retrieves previous candidates structurally similar to the current candidate and reuses their decisions and rationales as reference information for the current discrimination task. User confirmations or corrections and subsequent verification results are then used to revise the outcome labels, and validated cases are retained in memory. Precedents serve only as examples for comparison by the discrimination agent, not as ground-truth decisions that directly replace judgment for the current part. Retrieval depends on the candidate's geometric structure within the part rather than textual similarity between requests. Retrieved precedents are included in the discrimination input as a reference-only block. Absolute numerical values from previous parts are masked to prevent the direct transfer of dimensions between parts. The discrimination agent must make its own decisions based on the current part's serialized information and functional context rather than copy a precedent.

Extended candidate dimensions outside the predefined feature graph grammar are obtained through the measurement protocol. Numerical values for new dimensions are supplied only through the output of host-executed measurement code, rather than entered directly into the specification by the discrimination agent. Within a fixed attempt budget, the agent submits read-only measurement procedures for execution in an isolated host environment. Instead of explicitly entering coordinate values, the code identifies measurement targets using the entity identifiers assigned during serialization. The host resolves these identifiers to B-rep entities extracted by the kernel. Any value declared by the discrimination agent as the default for a new variable must agree with the host-recorded measurement within absolute and relative tolerances; otherwise, it is rejected.

This approach restricts the path through which the discrimination agent can introduce unmeasured numerical values, but it does not eliminate the possibility of selecting the wrong entity or writing an inappropriate measurement expression due to a misinterpretation of design intent. A successful measurement therefore does not automatically qualify the dimension as a design variable. The code that produced the measurement is retained and rerun unchanged at the \(\delta\)-following gate in Section~3.5. If an error in measurement-target selection or variable wiring prevents the candidate from passing the acceptance gate, the variable remains locked at \([d,d]\), and the failure reason and measurement code are recorded.

\subsection{Stage 2 (b): Design Logic Emission}
\label{sec:emission}

The second subprocess of Stage~2 converts the finalized discrimination results and kernel measurements into an executable Grasshopper definition. It does not infer new design intent; instead, it emits the preceding discrimination results and geometric feature graph as a canvas structure according to predefined rules.

Once discrimination is finalized, code generation becomes a translation based on predefined mappings. It is therefore not delegated to probabilistic generation. A rule-based emitter converts the feature graph assembled from kernel measurements into an executable Grasshopper definition, using discrimination results that have passed the consistency checks. Design variables (dv) are emitted as editable sliders, while design parameters (dp) are emitted as sliders locked to $[d,d]$. Locked values are also represented as sliders so that they can be tested through the same execution path as design variables during subsequent acceptance and verification. A dimension assigned \textsf{MIRROR} does not receive an independent slider; it is connected to the reference slider so that it follows changes in the reference value. However, if the measured values of the dependent and reference dimensions differ beyond tolerance, this connection could alter the default shape. The dependency declaration is therefore rejected and replaced with \textsf{FIXED}, and the reason is recorded in the audit log.

Dependency connections are stored in the structured feature graph before emission so that they persist when the Grasshopper definition is regenerated. The discrimination specification refers to dimensions through predefined variable identifiers, allowing the system to check the consistency of variable references without interpreting arbitrary names.

The emitter obtains numerical coordinates and dimensions from kernel measurements or approved coordinate expressions. Approximation enters not through these numerical values themselves but through the interpretation of which build operations should use them. For example, a cylindrical face spanning only part of an axial interval may represent either a protruding boss or the rounded end wall of a milled pocket, requiring different additive or subtractive operations. Because geometric classification alone may not resolve this ambiguity, the emitter evaluates material occupancy in candidate regions to determine the cutting direction and extent. Residual interpretation errors arising from the restricted build grammar are identified through regional metrics and localized diagnostics in the verification stage, which narrow the scope of subsequent corrections.

\subsubsection{Binding measurement-derived design variables}

The measurement protocol promotes new design variables that lie outside the predefined feature graph grammar, so their connections to the measured geometry must be defined explicitly. Delegating these connections to unrestricted LLM-generated code could reintroduce unverified numerical values and coordinates, despite the restrictions imposed during measurement. The discrimination agent is therefore allowed to declare only one of four predefined binding types: symmetric expansion of a span about its center, movement of only one endpoint of a span, making a single coordinate follow a slider, or adjustment of the body silhouette width. A rule-based binder translates the declaration into actual coordinate expressions.

The resulting expressions are constructed to recover the original feature graph coordinates before binding when the slider is at its default value. This is a condition at the coordinate-expression level; preservation of the final solid's default shape is verified separately through the acceptance gate in Section~3.5. A new slider is initially created in the locked state $[d,d]$, and its range is opened only after it passes the acceptance gate.

\subsubsection{Deterministic fillet matching}

Edge indices for filleting depend on the topology of a particular B-rep. Predetermining these indices from serialized STEP information can therefore select different edges after base-shape reconstruction or parameter editing. We first build the base shape without fillets and then perform edge selection and fillet application on the current geometry during subsequent updates. In the source STEP model, smooth AAG adjacency relations identify candidate cylindrical fillet faces, and the analytic parameters of the two neighboring faces on either side of each fillet are extracted. In the live B-rep, matching is attempted only when both neighboring faces are present. The selected edges must be parallel to the fillet axis, have midpoints within an allowed neighborhood of the feature center, and lie within the axial span. If filleting the complete edge set fails, smaller edge sets are retried within a fixed budget. If all attempts fail, the input B-rep is returned unchanged so that a fillet failure does not invalidate the already verified base shape.

\subsubsection{Kernel derivation of validity constraints and injection of live clamps}

The decision to exclude validity constraints $C_{\mathrm{val}}$ from the discrimination output in Eq.~\eqref{eq:discrimination} and derive them through the kernel is based on empirical observations. LLM-generated validity couplings may include unsupported relations, such as containment between unrelated bores. Determining whether two dimensions can geometrically interfere is a measurement problem rather than a matter of design-intent interpretation and is therefore assigned to the kernel under the role-separation principle.

The kernel detects dimension pairs that can be geometrically coupled in the feature graph and represents them as triples $(\theta_{\mathrm{dep}},\,\mathrm{rel},\,\theta_{\mathrm{anchor}})$ using the closed relation vocabulary
$\{\textsf{INSIDE},\textsf{WALL},\textsf{NEIGHBOR},\textsf{DEPTH},\textsf{OVERLAP}\}$.
Here, $\theta_{\mathrm{dep}}$ is the dependent dimension whose range is restricted, and $\theta_{\mathrm{anchor}}$ is the reference dimension determining its limit. Although the limit computation differs by relation, the common principle is to measure the geometric boundary at which a validity predicate changes and then shift the limit inward into the valid region by a kernel margin $\varepsilon_g$, avoiding boundary configurations at which Boolean operations may become unstable.

\textsf{INSIDE}, \textsf{WALL}, \textsf{NEIGHBOR}, and \textsf{DEPTH} impose upper bounds, clamping the dependent dimension to the minimum of its value and the limit. \textsf{OVERLAP} imposes a lower bound, taking the maximum instead. For \textsf{INSIDE} and \textsf{WALL}, the upper bound is based on a reference radius or the shortest distance to the body silhouette. For \textsf{DEPTH}, the boundary is the axial thickness of a tier or stack that the cut must not penetrate. When $n$ identical features are arranged on a circle of pattern radius $R_{\mathrm{pcd}}$, the feature radius at which adjacent features touch is calculated as $R_{\mathrm{pcd}}\sin(\pi/n)$.

For an attached boss whose center lies outside the body silhouette, \textsf{OVERLAP} uses the distance from the boss center to the silhouette boundary as a lower bound so that the overlap between the boss and body does not disappear. Relations are emitted only when their computed limits can become active within the declared slider ranges. Constraint injection therefore narrows the editing ranges toward geometrically feasible configurations without expanding the existing ranges.

The injection method depends on the state of the reference dimension. If the reference is a locked design parameter, the limit is written statically into the dependent slider's range. If the reference is a design variable, the limit varies with its value. In this case, the canvas computes the relation-specific limit and clamps the dependent input accordingly. This value clamp protects the geometry during every solve, while the dependent slider's displayed range is updated from the same constraint expression. The former restricts the actual input to geometry construction, and the latter restricts the range available to the user, providing geometric protection and interface feedback, respectively. Immediately after clamp injection, the volume at $\theta_0$ is compared with its pre-injection value to check that an incorrectly computed limit has not altered the already verified default shape.

\subsection{Stage 3: Design Verification --- Checking Default-Shape Preservation and Edit Validity}
\label{sec:verification}

The verification stage evaluates condition~(i), default-shape preservation, and condition~(iii), edit validity, defined in Section~3.1, together with the validity of new design variables promoted through the measurement protocol, using actual build results. This stage maintains the separation of roles. The kernel-based Verifier produces reproducible measurements and validity predicates from the geometric kernel and canvas state. At the same time, the verification agent handles the open-ended judgment required to interpret discrepancies and determine subsequent corrections. The agent does not write unrestricted geometry code; it proposes instructions only within a predefined correction scope. The loop for improving default-shape fidelity continues until the combined MATCH criterion is satisfied or the retry budget is exhausted. The attempt with the highest global IoU is retained and then undergoes the acceptance gate for measurement-derived variables and edit-validity verification. The system stores the generated definition and verification results from each attempt separately.

When verification reveals a discrepancy, the system returns diagnostic information to the design-logic construction process in Stage~2. This feedback does not require the LLM to regenerate the entire Grasshopper program. Instead, it expresses specific errors identified in the verification report as structured correction instructions. It reapplies them to the existing feature graph and deterministic Emitter.

\subsubsection{Empirical evaluation of default-shape preservation}

Because the rebuilt STEP solid $\widehat S=G(\theta_0)$ and source STEP solid $S$ may follow different global placement conventions, shape fidelity is evaluated in a part-local frame. The axis orientations of both shapes are retained, their bounding-box centers are aligned, and volumetric overlap is then computed. The IoU values below therefore measure local shape fidelity after removing global translation and do not indicate preservation of the absolute position in the source STEP file. The report also records the applied alignment translation.

Global $\operatorname{IoU}_{3D}$ is the larger of the two estimates. Boolean IoU uses the intersection volume computed by OpenCASCADE. Grid IoU partitions the combined bounding box of the two shapes into a grid. It uses cell-center samples to determine whether each point lies inside the source and rebuilt shapes. Both estimates are used because surfaces in Grasshopper-exported STEP files may be re-expressed as B-splines or become nearly coplanar with source surfaces, causing Boolean intersection to underestimate overlap or fail even when the shapes are nearly identical. Grid IoU supplements Boolean computation in these cases rather than replacing it. Because selecting the maximum may yield an optimistic fidelity estimate, both values are reported in Table~\ref{tab:shape_preservation}. Operation failures are distinguished from valid IoU values.

Global IoU alone is insufficient to assess the target. Excess volume from a body-silhouette approximation may obscure interface-region accuracy, while omitting a small hole may be masked by the volume of a large body. We therefore divide source faces into analytic-feature regions comprising holes, bosses, planes, and conical faces, and the remaining irregular body regions. Each grid point is assigned to the region of its nearest source face. In analytic-feature regions, IoU is computed using only the grid points assigned to those regions and denoted by $\operatorname{IoU}_{\mathrm{af}}$. The proportion of grid points where the rebuilt shape extends outside the source is denoted by $e_{\mathrm{breach}}$. In irregular body regions, only excess reconstruction is restricted using the rebuilt-to-source occupied-volume ratio $r_{\mathrm{org}}$. Under-reconstruction also affects global IoU, but a small local omission may still pass the global criterion; therefore, global IoU does not indicate detection of every local defect. This partition is a deterministic classification based on surface type, not a determination of face function.

Hole features are compared by the number detected at each radius, and $n_{\mathrm{missing}}$ counts holes detected in the source but not confirmed in the rebuilt model. Because cylindrical walls may be re-expressed as B-splines in the rebuilt STEP file, cylinders are refitted, and large cylindrical faces on the part exterior are excluded from hole counts. Cylinders detected only in the rebuilt model are retained as diagnostics because they may result from incorrect B-spline fitting; they are not currently included in the MATCH failure conditions. Thus, the missing-hole condition in Eq.~\eqref{eq:default-match} checks radius-wise count differences and does not establish correspondence in the position, axis, and depth of every individual hole.

Beyond the pass/fail decision, the Verifier identifies discrepancy locations and types. Each X, Y, and Z axis is divided into an equal number of bands. A band is marked as a hotspot if its volume difference between the rebuilt and source shapes exceeds a threshold fraction $\beta$ of the larger band volume. Differences in bounding-box extent along each axis are recorded to check that center alignment does not conceal size differences. For the symmetric Chamfer Distance, both solids are sampled using identical tessellation settings, and surface point clouds are formed from triangle vertices and centroids. Nearest-neighbor distances are computed in both directions, from the rebuilt shape to the source and from the source to the rebuilt shape. The mean of the two directional means, the 95th percentile, the maximum, and values normalized by the bounding-box diagonal are reported. These values diagnose local defects and are not directly used in MATCH. Finally, overlaid orthogonal projections of the source and rebuilt models support spatial interpretation of band hotspots and feature differences.

\subsubsection{Multilevel empirical evaluation of edit validity}

Edit validity is evaluated by applying the five validity predicates defined in Eq.~\eqref{eq:validity} in Section~3.1 to actual Grasshopper build results.

These predicates are evaluated at three sampling levels. First, a single-axis sweep moves one design variable from its default value toward its lower and upper bounds in $k$ steps in each direction. In contrast, all other variables remain at $\theta_0$. If a failure is observed in either direction, a bounded bisection search is performed between the last valid sample and the first failed sample. Equation~\eqref{eq:validity} is reapplied at each midpoint, and the valid--invalid interval is narrowed until a predefined iteration limit or the slider's value resolution is reached. Search values are quantized according to the variable-specific step convention toward the valid region on the default-value side, and the last confirmed valid value becomes the released range bound in that direction.

This boundary refinement procedure resolves the local boundary within the first observed valid--invalid interval. Given the numerical nonmonotonicity of Boolean operations, it does not claim to identify the global boundary of the entire valid region. Failed samples observed during boundary search are recorded separately as evidence for range adjustment and are excluded from the final edit-validity decision. In directions constrained by kernel-derived clamps, the sweep proceeds only to the geometric limit, followed by an additional test at the declared range endpoint.

After single-axis trimming, corner and interior configurations are newly sampled and evaluated within the final released ranges. Second, corner probes search for coupled failures caused by multiple sliders simultaneously reaching their range endpoints. When the number of degrees of freedom is small, all $2^n$ corners are tested; otherwise, pairs of variables are selected, and their four extreme combinations are tested. Third, interior probes use Latin hypercube sampling within the editing space and record each sample's validity, the overall valid-sample ratio, and the failure points. The current implementation records failures found by corner and interior probes as candidates for undeclared couplings, but it does not automatically convert them into new relational constraints or live clamps. Automatic adjustment of released ranges is therefore limited to boundaries directly identified by single-axis sweeps and boundary refinement.

Let $T_{\mathrm{axis}}^{\mathrm{final}}$ denote the single-axis samples retained within the final released ranges, including valid samples confirmed during boundary refinement and the last verified boundary points. These retained axis samples are not rerun after trimming. The corner and interior sample sets, $T_{\mathrm{vertex}}^{\mathrm{final}}$ and $T_{\mathrm{int}}^{\mathrm{final}}$, are newly generated and evaluated within the post-trim ranges. Define $T=T_{\mathrm{axis}}^{\mathrm{final}}\cup T_{\mathrm{vertex}}^{\mathrm{final}}\cup T_{\mathrm{int}}^{\mathrm{final}}$, retaining both passing and failing corner and interior configurations. The final empirical edit-validity decision in Eq.~\eqref{eq:edit-valid} therefore includes the newly tested multivariable configurations and is restricted to the tested samples.

\begin{equation}
\mathrm{editValid}(G)=\bigwedge_{\theta\in T}\mathrm{valid}(G(\theta)).
\label{eq:edit-valid}
\end{equation}

Thus, $\mathrm{editValid}(G)=1$ does not imply validity throughout the continuous editing space. It indicates only that no failures were observed among the single-axis and multivariable samples actually tested in this study.

\subsubsection{Acceptance gate for measurement-derived design variables}

Design variables promoted through the measurement protocol originate outside the predefined feature graph grammar. They are therefore not admitted as design variable sliders solely on the basis of successful measurement. A new variable $q$, initially locked at $[d,d]$, must satisfy the following three-part acceptance condition.

\begin{equation}
\begin{gathered}
\mathrm{accept}(q)\;\Longleftrightarrow\;\cnum{1}\land\cnum{2}\land\cnum{3},\\
\cnum{1}:\; \lvert V(G(\theta_0))-V_{\mathrm{ref},q}\rvert/V_{\mathrm{ref},q}\le\tau_V,\\
\cnum{2}:\; \bigl\lvert m_q\bigl(G(\theta_0+\delta e_q)\bigr)-(d_q+\delta)\bigr\rvert\le\kappa_m\tau_m,\\
\cnum{3}:\; \textstyle\bigwedge_{\theta\in T_q}\mathrm{valid}(G(\theta)).
\end{gathered}
\label{eq:dof-accept}
\end{equation}

First, the $\theta_0$-preservation gate in Eq.~\eqref{eq:dof-accept} checks whether the volume of the rebuilt model with the new variable set to its measured default $d$ agrees with the volume $V_{\mathrm{ref},q}$ of the reference build before variable binding within a relative tolerance. This provides an initial check for substantial changes to the default build. Although the coordinate expressions are constructed to recover the original coordinates at the default value, equal volume alone does not guarantee geometric identity of the final solid. Final default-shape preservation is therefore checked separately using the combined criterion in Eq.~\eqref{eq:default-match}.

Second, the $\delta$-following gate checks whether the new slider actually controls the intended geometric dimension. The slider value is changed from its measured default $d_q$ to $d_q+\delta$, and the geometry is regenerated and exported as STEP. The frozen measurement script $m_q$ originally used to measure variable $q$ is then applied unchanged. If the new measurement agrees with the target $d_q+\delta$ within tolerance, the slider change is considered to have been correctly reflected in the intended dimension.

Here, $\kappa_m$ is a measurement-tolerance factor accounting for numerical variations during STEP re-export and remeasurement. Promoted variables join the common sweep of the resulting definition. Accordingly, $T_q$ in Eq.~\eqref{eq:dof-accept} is the common final verification set $T$: retained single-axis samples within the final ranges, together with corner and interior samples newly generated and evaluated in those ranges. Axis samples are not rerun after trimming, and failed corner or interior samples are not excluded. The third gate thus checks sampled validity of the resulting definition, rather than only collecting configurations already known to be valid for $q$ in isolation.

A measurement that returns a fixed value regardless of the geometry, or a binding that does not propagate dimensional changes, can be detected by the second term of Eq.~\eqref{eq:dof-accept}. However, an incorrectly selected measurement target may exhibit the same change. Passing this term alone therefore does not establish the functional appropriateness of the measurement target. In execution, conditions~\cnum{1} and~\cnum{2} are evaluated first as the acceptance gate to reduce computational cost and reuse existing modules. Only variables passing both are opened to provisional ranges around their measured values and then included in the common sweep for all design variables described in Section~3.5.2 to evaluate condition~\cnum{3}. Failed variables remain at $[d,d]$, and their failure reasons and measurement scripts are recorded.

\subsubsection{Verification-report-guided correction and retention of the best attempt}
The retry loop supplies the verification agent with the combined MATCH result and its component metrics, band hotspots, and hole-feature deltas. It also provides source and rebuilt STEP summaries generated by the same parser, the Grasshopper status report, output metadata, and view captures.

The verification agent interprets failure causes from measurements but does not generate unrestricted geometry code for subsequent builds. Automatic correction is limited to failure types observed in the reconstruction process. Missing circular holes, cylindrical recesses, rectangular slots, and pockets are represented by predefined subtractive operations. Incorrect cutting depths for features already represented in the graph, body truncation caused by a side profile, unnecessary modules, and incorrect module combinations are addressed through predefined depth, profile, and module-selection rules. The procedure does not cover missing additive features such as bosses or ribs, freeform-surface discrepancies, arbitrary changes to feature positions or axes, or complex topological errors outside the current feature graph. Verification results modify the feature graph only through a closed correction vocabulary and predefined field schema; outputs containing unsupported types or fields are not incorporated.

Additional cuts and correction instructions from the verification agent are parsed according to a predefined schema. Module-targeted corrections are restricted to entries that actually exist in the measured feature graph, and coordinates for additional cuts are taken from values confirmed by STEP difference analysis. An additional cut is accepted only after deterministic checks confirm that it is subtractive, that its coordinates and cutting interval lie within an allowed region around the part bounding box, and that it does not remove the entire part. For STEP inputs, only validated additional cuts and restricted correction instructions are applied to the existing kernel-generated measured feature graph. The Emitter then deterministically regenerates the Grasshopper definition.

Because shape fidelity does not necessarily improve monotonically across retries, the generated definition and verification results from each attempt are stored individually. The attempt with the highest global IoU is retained, and edit validity is evaluated only for that attempt. Its verification status combines the MATCH criterion and the measured edit-validity result.

\begin{equation}
\mathrm{VALID\_RESULT}
=
\mathrm{MATCH}_{0}(\widehat S,S)
\land
\mathrm{editValid}(G).
\label{eq:valid-result}
\end{equation}

If the retained highest-IoU attempt satisfies \textsf{VALID\_RESULT}, it is reported as a technically verified result. Otherwise, it is retained as an \textsf{UNVERIFIED\_CANDIDATE}, rather than reported as a valid conversion. Edit validity is not evaluated for the other attempts. This separates IoU-based candidate retention from the subsequent determination of conversion success.

The \textsf{VALID\_RESULT} status in Eq.~\eqref{eq:valid-result} denotes technical verification of both default-shape preservation and sample-based edit validity. Because condition~(ii), design-intent reflection, cannot be determined automatically through geometric verification, it is excluded from this status and evaluated separately in the experiments.

\subsection{Stage 4: Product Assembly}
\label{sec:assembly}

Stage 4 takes the verified part definitions and produces a single canvas where the parts are placed, their shared dimensions are linked, and the released ranges of user-facing sliders are assessed for assembly compatibility. The separation of roles is the same as in the part-level stages: the kernel measures, the LLM makes one discrete judgment, and the kernel executes and verifies it. No part-specific rule or expected pose is supplied.

\subsubsection{Joint candidates and confirmation}
For every pair of parts, the kernel lists the contacts that could be a joint: a pin candidate is a bore of one part and a bore or boss of another with matching diameters; a bolted candidate is a hole pattern of one part and a pattern of another with the same count, diameter, and pairwise distance matrix. Each candidate carries its measured evidence and the hole-to-hole correspondence. The Relate LLM agent, the third agent in the framework, receives this candidate list and returns the subset of candidates that form the assembly, along with each joint type and a rationale. It cannot add a contact the kernel did not measure and specifies no position. Confirmed joints form a graph over the parts; a cycle of pin joints is a closed kinematic loop, and a rigid joint attaches a child part to a parent.

\subsubsection{Pose inference}
\label{sec:assembly-pose}
The kernel derives the pose of every part from the joint graph. A closed loop is solved as a planar linkage with one part as ground and the measured pin-to-pin distances as link lengths. Both elbow configurations and link reversals are enumerated. Candidate configurations are ranked by interference, the proportion of closed configurations in a $\pm60^{\circ}$ sweep, and convexity, without using the source assembly as an inference input. The kernel places a rigidly attached part by its bolt pattern: it aligns the two hole sets with a two-dimensional Procrustes fit \cite{schonemann1966generalized} and evaluates the resulting symmetry-admissible poses. The kernel retains the interference-free pose with the largest contact area with the parent and records discarded alternatives.

\subsubsection{Product-level verification and repair}
The kernel checks the placed rebuilt parts for interpenetration and identifies reconstruction-induced intrusions. The kernel addresses identified intrusions with bounded subtractive corrections, after which it rebuilds the affected part once. The source assembly is used only as a separate evaluation reference and does not guide these corrections. The same pass records every seated contact, the measured thickness of the inner part, the slot width, their clearance, and the contact axis.

\subsubsection{Measured seat links} \label{sec:assembly-links}
A seated contact introduces a dependency between two parts: when the inner thickness changes, the receiving slot must follow. The kernel perturbs the outer part's sliders. It measures the resulting motion of the two slot faces along the contact axis, forming a Jacobian of face positions with respect to the sliders. For a thickness change $\Delta t$, the target face displacements are $(-\tfrac12,+\tfrac12)\Delta t$, preserving the initial clearance. Candidate subsets of one or two sliders are enumerated, and least-squares solutions meeting the residual tolerance are ranked first by the total absolute slider displacement per unit change in thickness $t$, and then by subset size. The solver compensates for other free sliders that move the faces and includes relative translation along the contact axis when permitted by the attachment. Each accepted solution is emitted as a linear dependency replacing the corresponding slider; these relations are not relinearized during editing. If the available controls cannot span the required face motion, the allowed increase in inner thickness is limited by the measured clearance, and the missing degree of freedom is recorded.

\subsubsection{Canvas integration}
The part definitions are merged into one canvas with the relations above represented explicitly: shared sliders for dimensions marked as shared by discrimination and measured as equal by the kernel; a live solution of each closed loop with the input angle clamped to the closable range; an updated joint reference for each attached part, so that a slider moving a joint hole on either side moves the attached part with it; and a lock on any slider that would move a bolted hole the coupled part cannot follow. User-facing sliders are collected in one control panel.

\subsubsection{Product-level dynamic clamping}
\label{sec:assembly-sweep}
Each panel slider is swept from its default toward both bounds. At each step, the parts it touches are rebuilt and placed with the transforms the canvas graph delivers, and the implemented product test $\bigl(\bigwedge_i\mathrm{built}_i(\theta)\bigr)\land\mathrm{assembled}(G(\theta))$ is applied, using the assembly conditions in Eq.~\ref{eq:validity-product}: loop closure, interpenetration depth between every pair of placed parts, and the integrity of every confirmed bolted joint. The first failed step is refined by bisection, the slider's range is rewritten to the last valid value, and the reason is recorded. The sweep is one-dimensional by default, as at the part level.

\section{Experiments and Results} \label{sec:experiments}
This section evaluates the proposed framework against the three conditions defined in Section ~\ref{sec:problem}: default-shape preservation, design-intent reflection, and edit validity. The experiments address three questions. First, to determine whether the generated Grasshopper definition reproduces the input STEP geometry when all sliders are set to their default values. Secondly, the objective is to ascertain the dimensions selected by the discrimination LLM as design variables from the serialized geometric information, and the manner in which it establishes dependencies. In addition, the implementation of the selected dimensions as actual editing variables is traced. Third, representative assembly-level examples are used to examine whether the reconstructed definitions support coordinated changes in component geometry and assembly configuration. These examples illustrate the geometric variations obtained through parameter adjustments, with checks for geometry regeneration, mechanism closure, joint preservation, and inter-part interference. Furthermore, an analysis of changes in variable selection in response to user requests and the contributions of serialization channels is conducted.

The reconstruction of shapes, the selection of dimensions, the implementation of variables, and the successful dimensional editing are regarded as discrete evaluation stages. High geometric similarity alone does not constitute evidence of design-intent reflection. Furthermore, the appearance of a requested dimension in the discrimination output is not counted as successful editing unless its implementation through actual connections and remeasurement is confirmed. The product-level editing examples illustrate coordinated changes in component geometry and assembly configuration through shared parameters and measured dimensional dependencies.

\subsection{Experimental Setup}
\label{sec:setup}

\subsubsection{Evaluation parts}
The evaluation uses five parts from an actual robotic arm design \cite{lee2025multi}: rocker, crank, connecting rod, frame, and joint upper. All inputs are STEP-format B-reps without design history, and each input loads as a valid single solid in OpenCASCADE. The connecting rod is a bar-shaped part connecting bearing holes at its two ends. At the same time, the crank contains multiple levels of cylindrical bosses. The frame includes repeated arms and bolt-hole patterns, and the rocker has a bifurcated fork structure. The joint upper consists of a large through-bore and groups of mounting holes at both ends. The product-level evaluation adds a sixth part, the joint holder, a bracket bolted to the rocker face that carries the joint upper, and treats the six parts as one product: the holder bolts to the rocker, and the joint upper bolts to the holder. The source assembly STEP of the arm is used only as an evaluation reference. It is not supplied as an input to any inference stage. The holder geometry was simplified from the original design for this evaluation, with its interface features retained. Like the other five parts, the framework first converted the holder at the part level, and the product-level evaluation uses the resulting part definitions.

Figure~\ref{fig:complexity} summarizes the complexity of the five robotic-arm parts used in the application, in terms of B-rep face count, face composition, and the number of entries in the kernel feature graph. All five parts consist solely of analytic surfaces. Hence, the evaluation focuses on dimension selection and dependency recovery rather than on free-form surface handling.

\begin{center}
\includegraphics[width=0.9\textwidth]{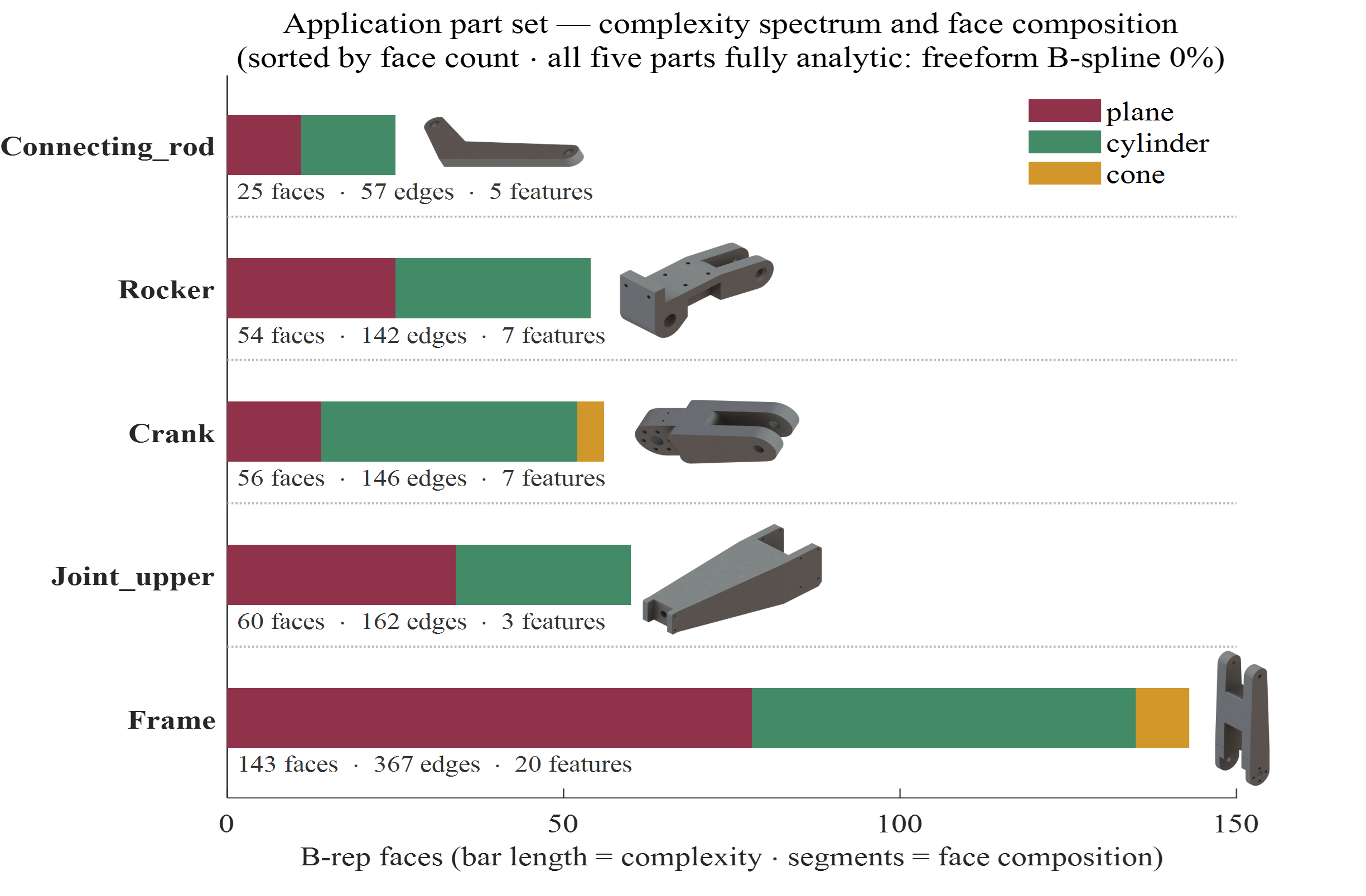}
\captionof{figure}{Complexity of the five evaluation parts, sorted by B-rep face count. Bar length is the number of faces; the segments show the face composition (plane, cylinder, cone). The annotation under each bar gives the number of B-rep faces, B-rep edges, and entries in the kernel feature graph. All five parts are fully analytic (no free-form faces).}
\label{fig:complexity}
\end{center}

These cases include different editing targets: bore radius, inter-axis distances, bolt patterns, and overall dimensions, as well as fork gaps, wall thicknesses, boss steps, and hole-group positions. The former are often already represented as parameters in the kernel feature graph, while the latter require additional measurement and connections to become editable variables. A dimension outside the feature graph does not mean a dimension that the geometric kernel cannot measure; it means that the initial feature graph lacks a variable that directly controls that dimension.

Because the evaluation geometries primarily comprise analytic surfaces, the experiments mainly address dimension selection and the feasibility of parametrization for mechanical parts. However, analytic surfaces do not guarantee that the serialized relations are complete or that measurement targets are interpreted correctly. In particular, the spacing between inclined faces and the positions of symmetric features can correspond to different dimensions depending on the measurement direction and target definition. We therefore distinguish errors in geometric representation, target interpretation, variable connections, and edit verification.

\subsubsection{Implementation environment}

Claude Sonnet 4.6 was used for LLM-based discrimination, variable promotion, and verification-report interpretation. The reasoning effort was set to low for discrimination and promotion and to medium for verification-report interpretation. Because the interface used does not support a seed, the main experiments were repeated five times per part under identical conditions to evaluate run-to-run variation in discrimination results and conversion success. OpenCASCADE 7.7.2 was used for geometric measurement and verification, while Rhinoceros 3D 8 and its integrated Grasshopper environment were used to generate and execute the parametric definitions. Canvas construction, parameter adjustment, and STEP export are automated within the Grasshopper environment. This study conducted baseline discrimination and request comparisons without precedent memory.

\subsubsection{Thresholds and sampling settings}

For the combined shape-preservation criterion, the global and analytic-feature IoU thresholds were both set to 0.80, and the upper bounds on the irregular-body volume ratio and boundary-breach ratio were set to 3.0 and 0.15, respectively. Edit verification required the generated volume to exceed 2\% of the default volume. Single-axis sweeps tested six steps in each direction from the default value. In contrast, multivariable tests used up to 80 corner configurations and 20 interior samples generated by Latin hypercube sampling.

For acceptance testing of promoted variables, the increment used to test change following was 4~mm, the tolerance for agreement with the default measurement was 1.0~mm, and the provisional editing range was set to $\pm40\%$ of the measured value. Coaxial merging in the AAG used a radius-difference tolerance of 0.05~mm, an absolute axis-direction dot product greater than 0.999, and an inter-axis distance tolerance of 0.2~mm. The angular threshold for tangential continuity was $8^{\circ}$.

\subsection{Evaluation Protocol and Metrics}
\label{sec:metrics}

\subsubsection{Default-shape preservation}

Default-shape preservation is evaluated using a combined criterion incorporating global $\mathrm{IoU}_{3D}$, analytic-feature-region $\mathrm{IoU}_{\mathrm{af}}$, the irregular-body volume ratio, the boundary-breach ratio, and missing features. The metrics are computed as described in Section~3.5. Symmetric Chamfer distance, differences in bounding-box dimensions, and additionally detected features are also reported as auxiliary metrics. Passing the combined criterion indicates that the specified thresholds are satisfied, not that the source and rebuilt geometries are strictly identical.

\subsubsection{Design-intent reflection}

Design-intent reflection is analyzed through the structure of discrimination results, run-to-run consistency, the promotion process, and responsiveness to requests. Dimensions declared as OPENED or MIRROR are distinguished from editing variables that are actually accepted. We record the numbers of selections inside and outside the feature graph and whether dependency declarations are implemented. Run-to-run consistency is measured by the Jaccard similarity between selection sets A and B and reported as the mean over the 10 possible pairs among five runs for each part. Responsiveness to a user request is evaluated by whether discrimination selects the requested dimension as an independent variable; merely citing the request in a natural-language rationale is not considered fulfillment. At the product level, editing examples illustrate how confirmed assembly relations and cross-part dimensional couplings support coordinated changes in the assembly. The examples show the role of shared variables and measured links in maintaining compatibility between joined parts.

\subsubsection{Assembly-level editing and representative cases}

The exposed design parameters were adjusted on the merged Grasshopper canvas to generate variations in component geometry and assembly configuration. Shared parameters and measured dimensional dependencies coordinated changes between connected components.

Each control-panel slider was swept from its default value toward both declared bounds. At each step, the affected parts were rebuilt and placed, and the implemented test $\bigl(\bigwedge_i\mathrm{built}_i(\theta)\bigr)\land\mathrm{assembled}(G(\theta))$ was applied, with assembly conditions defined in Eq.~\eqref{eq:validity-product}. The checks covered geometry
regeneration, mechanism closure, preservation of confirmed bolted joints, and inter-part penetration relative to the tolerance $\tau_d$. When a check failed, the boundary was refined by
bisection, and the corresponding range bound was set to the last verified valid value. Range adjustments and reasons for locking sliders were recorded. The checks were performed on the base
geometry without fillets.

For visualization, the run with the largest final number of design variables among those passing the shape-preservation criterion was selected as the representative case; ties were resolved by
selecting the run with the higher global IoU. For the default-shape comparison, run~1 was used because the shape-preservation metrics were identical across runs at the reported precision.
Shape-preservation pass rates and parameter-selection consistency were calculated using all runs in the corresponding experimental group.

\subsection{Default-Shape Preservation Results}
\label{sec:shape-results}

Table~\ref{tab:shape_preservation} presents the combined criterion results and their component metrics for each part, and Figure~\ref{fig:gt_rebuilt} shows the source and rebuilt models side by side. Figure~\ref{fig:rocker_iou} illustrates how the Boolean IoU is obtained for the rocker: after bounding-box-center alignment, the kernel computes the common solid $A\cap B$ and the two difference solids, $A\setminus B$ (material present only in the source) and $B\setminus A$ (material present only in the rebuilt model), and $\mathrm{IoU}_{\mathrm{bool}}=V(A\cap B)/V(A\cup B)$. For the rocker the residual volume is confined to the fork-side end faces and the base of the prongs, which accounts for its $\mathrm{IoU}_{\mathrm{bool}}$ of 0.970. All 25 runs passed the combined criterion, with global IoU ranging from 0.977 to 0.997.

\begin{center}
\includegraphics[width=\textwidth]{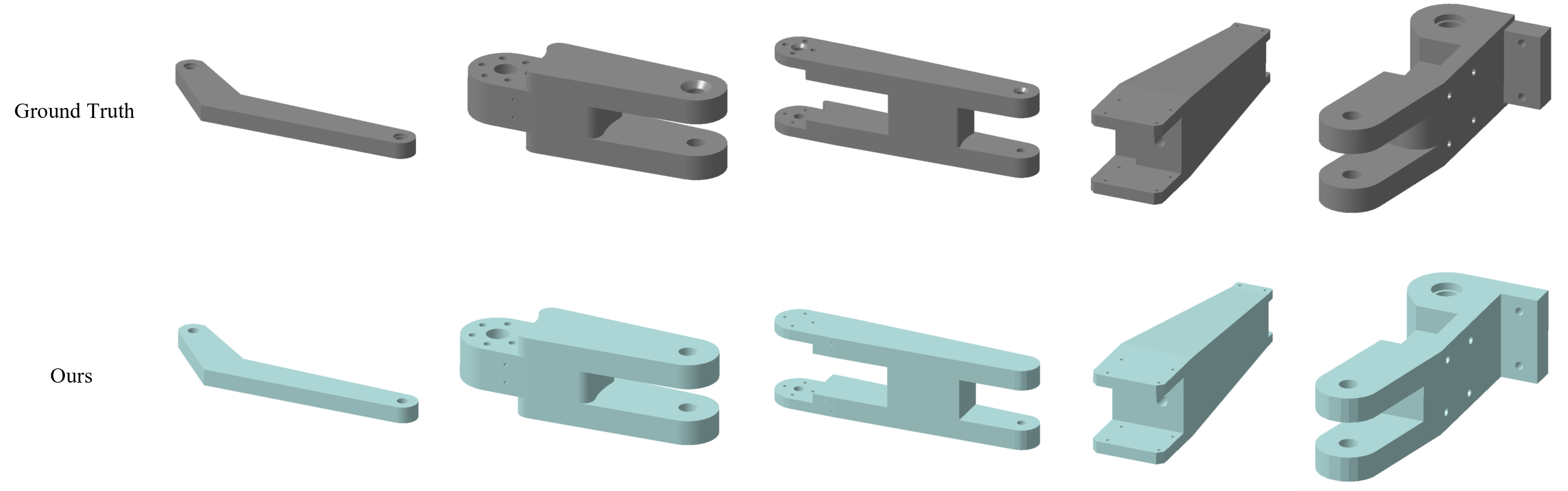}
\captionof{figure}{Source models (top) and rebuilt models at the default slider values (bottom) for the five evaluation parts: connecting rod, crank, frame, joint upper, and rocker. The rebuilt models are the run-1 outputs whose metrics are reported in Table~\ref{tab:shape_preservation}.}
\label{fig:gt_rebuilt}
\end{center}

\begin{center}
\includegraphics[width=0.5\textwidth]{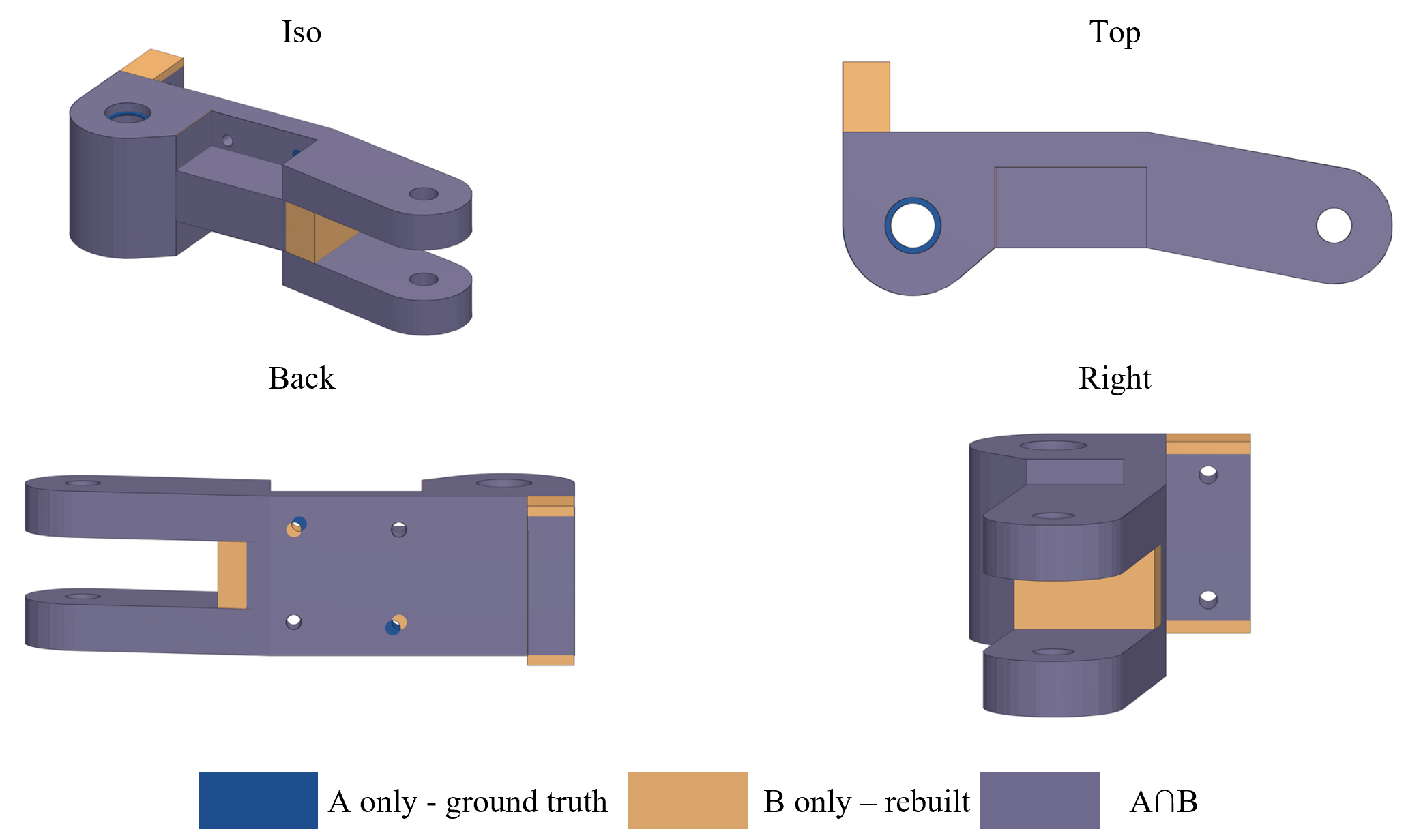}
\captionof{figure}{Boolean comparison of the rocker (source $A$ vs.\ rebuilt $B$) in four views. Grey-violet: common solid $A\cap B$; blue: $A\setminus B$ (present only in the source); tan: $B\setminus A$ (present only in the rebuilt model). $\mathrm{IoU}_{\mathrm{bool}}=0.970$ (Table~\ref{tab:shape_preservation}).}
\label{fig:rocker_iou}
\end{center}

For each part, IoU did not vary across runs at the precision reported in the table. Analytic-feature-region IoU differed from global IoU by at most 0.004, the boundary-breach ratio was well below its upper bound of 0.15, and the missing-hole count was zero for all five parts.

\begin{table}[htbp]
\centering
\caption{Default-shape preservation results by part.}
\label{tab:shape_preservation}
\begingroup
\small
\renewcommand{\arraystretch}{1.25}
\setlength{\tabcolsep}{4pt}
\begin{tabular*}{\linewidth}{@{\extracolsep{\fill}}lccccc@{}}
\toprule
\multirow{2}{*}{Part}
& \multirow{2}{*}{\makecell{Combined\\passes}}
& \multicolumn{4}{c}{Shape similarity} \\
\cmidrule(lr){3-6}
& & \makecell{Global\\$\mathrm{IoU}_{3D}$}
  & \makecell{Boolean\\$\mathrm{IoU}_{\mathrm{bool}}$}
  & \makecell{Grid\\$\mathrm{IoU}_{\mathrm{vox}}$}
  & \makecell{Analytic features\\$\mathrm{IoU}_{\mathrm{af}}$} \\
\midrule
Connecting rod & 5/5 & 0.997 & 0.997 & 0.997 & 0.997 \\
Crank          & 5/5 & 0.981 & 0.981 & 0.981 & 0.981 \\
Frame          & 5/5 & 0.987 & 0.980 & 0.987 & 0.987 \\
Rocker         & 5/5 & 0.977 & 0.970 & 0.977 & 0.977 \\
Joint upper    & 5/5 & 0.979 & 0.979 & 0.975 & 0.975 \\
\bottomrule
\end{tabular*}
\par\vspace{5pt}
\begin{minipage}{\linewidth}
\footnotesize\raggedright
\textit{Note:} Pass counts are based on five runs per part; all five runs produced identical metrics at the reported precision, and the values shown are from run~1.
The combined criterion follows Eq.~\eqref{eq:default-match}, whose two IoU terms are $\mathrm{IoU}_{3D}$ and $\mathrm{IoU}_{\mathrm{af}}$;
both must be at least 0.80, the boundary-breach ratio must not exceed 0.15, and the missing-hole count must be zero.
$\mathrm{IoU}_{\mathrm{bool}}$ is the kernel-computed volume ratio $V(A\cap B)/V(A\cup B)$ after bounding-box-center alignment;
$\mathrm{IoU}_{\mathrm{vox}}$ is the same ratio on a shared lattice with 48 cells along the longest axis;
$\mathrm{IoU}_{3D}=\max(\mathrm{IoU}_{\mathrm{bool}},\mathrm{IoU}_{\mathrm{vox}})$.
$\mathrm{IoU}_{\mathrm{af}}$ is evaluated on the analytic-feature region of the same lattice; because no irregular regions were present
($r_{\mathrm{org}}$ not applicable and its MATCH condition treated as true), it coincides with $\mathrm{IoU}_{\mathrm{vox}}$ for all parts.
\end{minipage}
\endgroup
\end{table}

Separating shape preservation from variable selection is important when interpreting these results. The discrimination LLM determines which dimensions remain editable, whereas the default shape is constructed from kernel measurements and the feature graph. Changes in the variable set therefore did not directly translate into changes in the default shape. The diagnostic metrics also showed no large discrepancies. Volume ratios ranged from 0.997 to 1.022, the maximum difference in bounding-box dimensions was 0.5~mm, mean Chamfer distances ranged from 0.33 to 1.06~mm, and no band hotspots exceeded the threshold. However, the repetition of identical metrics alone does not establish exact agreement of all local faces or recovery of design intent.

\subsection{Design-Intent Reflection Results}
\label{sec:intent-results}

\subsubsection{Discrimination structure and run-to-run consistency}

Table~\ref{tab:selection_consistency} summarizes the selection structure in the baseline discrimination experiment. The serialized dimension sets range from 18 dimensions for the connecting rod to 156 for the frame. The discrimination LLM selected an average of approximately 11--26 dimensions per part as OPENED or MIRROR. Derived dimensions determined by other dimensions, such as through-hole depth, were classified separately. MIRROR counts are a subset of total selections and are not added again to the other columns.

\begin{table}[htbp]
\centering
\caption{Selection structure and run-to-run consistency of baseline discrimination.}
\label{tab:selection_consistency}
\begingroup
\small
\renewcommand{\arraystretch}{1.25}
\setlength{\tabcolsep}{3pt}
\begin{tabular*}{\linewidth}{@{\extracolsep{\fill}}lccccccccc@{}}
\toprule
\multirow{2}{*}{Part}
& \multirow{2}{*}{\makecell{Input\\dims.}}
& \multicolumn{5}{c}{Selected dimensions}
& \multicolumn{3}{c}{Consistency ($J$)} \\
\cmidrule(lr){3-7}\cmidrule(lr){8-10}
& & Total & \makecell{In\\graph} & \makecell{Outside\\graph}
  & Derived & MIRROR
  & Total & \makecell{In\\graph} & \makecell{Outside\\graph} \\
\midrule
Connecting rod & 18  & 11.2 & 2.2 & 7.4  & 1.6 & 3.6 & 0.71 & 0.55 & 0.85 \\
Crank          & 123 & 20.0 & 8.4 & 11.6 & 0.0 & 2.6 & 0.69 & 0.93 & 0.54 \\
Frame          & 156 & 26.0 & 8.6 & 17.4 & 0.0 & 3.4 & 0.71 & 0.91 & 0.64 \\
Rocker         & 81  & 20.0 & 8.6 & 11.0 & 0.4 & 2.0 & 0.61 & 0.83 & 0.52 \\
Joint upper    & 24  & 19.2 & 4.0 & 12.2 & 3.0 & 2.0 & 0.61 & 1.00 & 0.49 \\
\bottomrule
\end{tabular*}
\par\vspace{5pt}
\begin{minipage}{\linewidth}
\footnotesize\raggedright
\textit{Note:} Selection counts are means over five runs per part.
``In/outside graph'' indicates whether a dimension belongs to the predefined feature graph grammar.
The total is the sum of in-graph, outside-graph, and derived dimensions.
MIRROR is a state-specific count included in the total and is not added separately.
$J$ is the mean Jaccard similarity over the 10 possible pairs of five runs.
\end{minipage}
\endgroup
\end{table}

\begin{table}[htbp]
\centering
\caption{Agreement and differences in repeated selection of targets outside the feature graph.}
\label{tab:selection_difference}
\begingroup
\small
\renewcommand{\arraystretch}{1.25}
\setlength{\tabcolsep}{4pt}
\begin{tabular*}{\linewidth}{@{\extracolsep{\fill}}lccccc@{}}
\toprule
\multirow{2}{*}{Part}
& \multirow{2}{*}{\makecell{Unique\\targets}}
& \multirow{2}{*}{\makecell{Identical\\declarations}}
& \multirow{2}{*}{\makecell{With\\differences}}
& \multicolumn{2}{c}{Difference type} \\
\cmidrule(lr){5-6}
& & & & \makecell{Representation\\difference} & \makecell{Selection\\difference} \\
\midrule
Connecting rod & 11 & 5  & 6  & 4 & 2  \\
Crank          & 19 & 5  & 14 & 6 & 8  \\
Frame          & 29 & 8  & 21 & 1 & 20 \\
Rocker         & 19 & 5  & 14 & 0 & 14 \\
Joint upper    & 27 & 10 & 17 & 0 & 17 \\
\bottomrule
\end{tabular*}
\par\vspace{5pt}
\begin{minipage}{\linewidth}
\footnotesize\raggedright
\textit{Note:} Unique targets are targets or intervals selected at least once across five runs, not a per-run mean.
Identical declarations indicate selection of the same target in all five runs using the same declaration set.
A representation difference indicates that the same target was selected in every run but with different declaration sets;
a selection difference indicates selection in only some runs.
``With differences'' is the sum of representation and selection differences.
\end{minipage}
\endgroup
\end{table}

For every part, selections outside the feature graph outnumbered those inside it. For the frame, in particular, an average of 17.4 of the 26 selected dimensions lay outside the feature graph. This indicates that discrimination goes beyond deciding whether to expose existing sliders and proposes additional editing targets. However, these proposed dimensions require subsequent measurement and connections before they can serve as actual control variables. Therefore, an increase in selection count cannot be interpreted as an increase in editing degrees of freedom.

For the four parts other than the connecting rod, Jaccard similarity for in-graph selections ranged from 0.83 to 1.00, while that for outside-graph selections ranged from 0.49 to 0.64. This indicates greater run-to-run variation in the declaration sets for dimensions outside the feature graph. However, this variation may reflect both changes in the selected targets and different representations of the same target, so similarity alone does not establish the stability of geometric target selection.

To distinguish the sources of disagreement, we compared referenced entities, measurement directions, and boundary locations in the source geometry. Radius and depth declarations spanning multiple features were expanded into individual feature targets. We then accounted for confirmed changes in reference order, differences in measurement type for the same face pair, inter-axis distances of coaxial features, and correspondences between identical axial intervals. After selecting these correspondences, we compared whether each target was selected in the five runs and which declaration sets identified it. In Table~\ref{tab:selection_difference}, ``Identical declarations'' indicates that all five runs selected the same target using the same declaration set. ``Representation difference'' indicates that the declaration sets differed but the same corresponding target was selected in every run. ``Selection difference'' indicates that the target was selected in only some runs even after accounting for correspondences. For the frame, rocker, and joint upper, most differences were selection differences rather than representation differences. Because this analysis is limited to confirmed geometric correspondences, the final column should not be interpreted as a definitive difference established after examining every possible physical equivalence.

\subsubsection{Promotion process}

Dimensions selected outside the feature graph are promoted to editing variables through the measurement protocol. Across five discrimination runs for each of the five parts, 237 promotion requests were submitted to the measurement protocol. Of these, 202 passed the measurement-agreement gate, and 108 were provisionally opened as editing variables. The measurement-gate pass rate was 85.2\%, and 53.5\% of candidates passing the gate were provisionally opened.

These counts use promotion requests submitted to the measurement protocol as the counting unit and therefore differ from the 298 outside-graph selections aggregated across all 25 runs in Table~\ref{tab:selection_consistency}. First, those selections comprise 251 \textsf{OPENED} declarations and 47 \textsf{MIRROR} declarations. The latter are connected to reference variables without creating independent sliders and are therefore not candidates for promotion. Second, the remaining 251 \textsf{OPENED} declarations do not correspond one-to-one to the 237 promotion requests. During measurement, the discrimination agent could combine declarations covering multiple features into a single request or split them into separate requests. Some declarations did not lead to measurement requests, while additional dimensions not included in the declarations were also requested.

The remaining 94 cases were categorized by the reason they were not opened. Of these, 18 were classified as through-hole depths, 14 were mapped to existing variables, and two were merged as duplicate declarations. These cases did not add new independent variables and were therefore not all counted as variable-implementation failures.

The 46 cases excluded at the binding stage included cases where no geometry-generating element controlling the target face could be found, the control target overlapped or conflicted with an existing variable, or the measured dimension did not match the binding target dimension. A further 14 cases were rejected because the remeasured change after modifying the variable did not agree with the specified change within tolerance. These results indicate the need to jointly check geometric control and the consistency of measurement and re-identification. Rejection alone, however, does not establish that the dimension is inherently uneditable.

The crank had the highest provisional opening rate at 0.81, and the connecting rod the lowest at 0.32. These differences may reflect part-specific differences in the dimensions selected by discrimination and the range of control variables available in the existing feature graph.

\subsubsection{Responsiveness to requests}
\label{sec:request}

The same geometry can be parameterized differently depending on which dimensions are chosen as driving dimensions. A two-pronged fork can be represented by ``lug thickness and overall height'' or by ``gap between the lugs and lug thickness.'' The default shape remains the same, but the sliders available to the designer differ. Without a request, discrimination leaves this choice to the LLM, so a dimension such as the gap accommodating the inserted part may remain dependent on other dimensions. The request experiment tests whether a designer-specified dimension is selected as an independent variable.

For each part, one target dimension was chosen from those selected zero or one time across the five runs without requests. Dimensions already adjustable through combinations of other dimensions were not excluded; only derived dimensions that could not be independent under any parametrization were excluded. The crank was excluded because the baseline discrimination LLM had already selected its key dimensions in all five runs, leaving no meaningful target satisfying this criterion. The experiment therefore covered the remaining four parts. Requests described only function, without identifiers or numerical values. For example, the request for the frame was: ``The large-hub side of this frame has a two-pronged fork, and the gap between its two inner faces should be adjustable during design.''

The request was appended to the end of the discrimination LLM input and, before discrimination, interpreted as a single dimension entry in the distance table. Evaluation counted the number of runs out of five in which the requested dimension was selected as an independent variable. A dependent assignment was not counted as independent selection. Merely citing the request in the natural-language rationale was not counted as selection.

All four target dimensions were present in the distance table, but without requests they were selected only once in 20 runs, as shown in Table~\ref{tab:request}. They were not selected because the discrimination LLM represented the same geometry using other dimension combinations. For example, without a request, the frame's fork was consistently parameterized with lug thickness and overall height as independent variables, leaving the gap between the lugs as a dependent value determined by their difference. With a request, the target dimension was selected as an independent variable in all five runs for the frame, rocker, and connecting rod, and in four runs for the joint upper (Table~\ref{tab:request}). These results measure selection responsiveness at the discrimination stage; actual variable connections and achievement of the requested dimensional change require separate execution-based verification.

\begin{table}[t]
  \centering
  \caption{Selection of requested dimensions as independent variables across five runs per part, with and without user requests.}
  \label{tab:request}
  \begin{tabular}{@{}l p{8.2cm} c c@{}}
    \toprule
    Part & Requested dimension (measured value) & No request & With request \\
    \midrule
    Frame          & Inner fork gap on the shoulder side (106 mm) & 0/5 & 5/5 \\
    Rocker         & Inner fork gap on the rod side (29 mm) & 1/5 & 5/5 \\
    Connecting rod & Wall thickness between counterbore bottoms at opposite ends (18 mm) & 0/5 & 5/5 \\
    Joint upper    & Internal height between upper and lower plates (62 mm) & 0/5 & 4/5 \\
    \bottomrule
  \end{tabular}
\end{table}

For the joint upper, multiple plane pairs could match the request description, and one run omitted the target pair. This suggests that ambiguity in interpreting the target may have contributed to the omission. The requested dimensions in Table~\ref{tab:request} include both void gaps in forks and the wall thickness of a material region in the connecting rod; they are therefore not all interpreted as empty-space dimensions.

These results show that requests did not introduce new dimensions but changed which of the dimensions already present in the input were treated as independent variables. Geometry alone does not determine which dimensions should be independent and which should be dependent. The need to directly control a fork gap arises from the assembly context, and the request provides a way to convey context absent from an individual part's serialized information to the discrimination LLM.

\subsection{Assembly-Level Parametric Editing}
\label{sec:editvalid-results}
The exposed design parameters of the reconstructed parametric assembly were adjusted to generate geometric variations while maintaining assembly constraints. Figure~\ref{fig:robot} presents 28 representative configurations of the robotic-arm assembly generated by adjusting the exposed design parameters, illustrating the range of geometric variations supported by the reconstructed parametric model. The resulting configurations exhibit variations in component geometry and mechanism configuration, demonstrating that the reconstructed model supports design modifications through parameter adjustments without requiring manual remodeling.

In the illustrated configurations, geometry regeneration was successful, mechanism closure and confirmed bolted-joint relationships were maintained, and inter-part penetration remained within the specified tolerance. In particular, geometric dependencies between connected components were maintained as the design parameters changed, allowing the assembly to adapt consistently rather than requiring independent modifications to individual parts. These results demonstrate the ability of the proposed framework to support coordinated parametric editing while preserving assembly-level geometric constraints.

\begin{center}
\includegraphics[width=\linewidth,height=0.7\textheight,keepaspectratio]{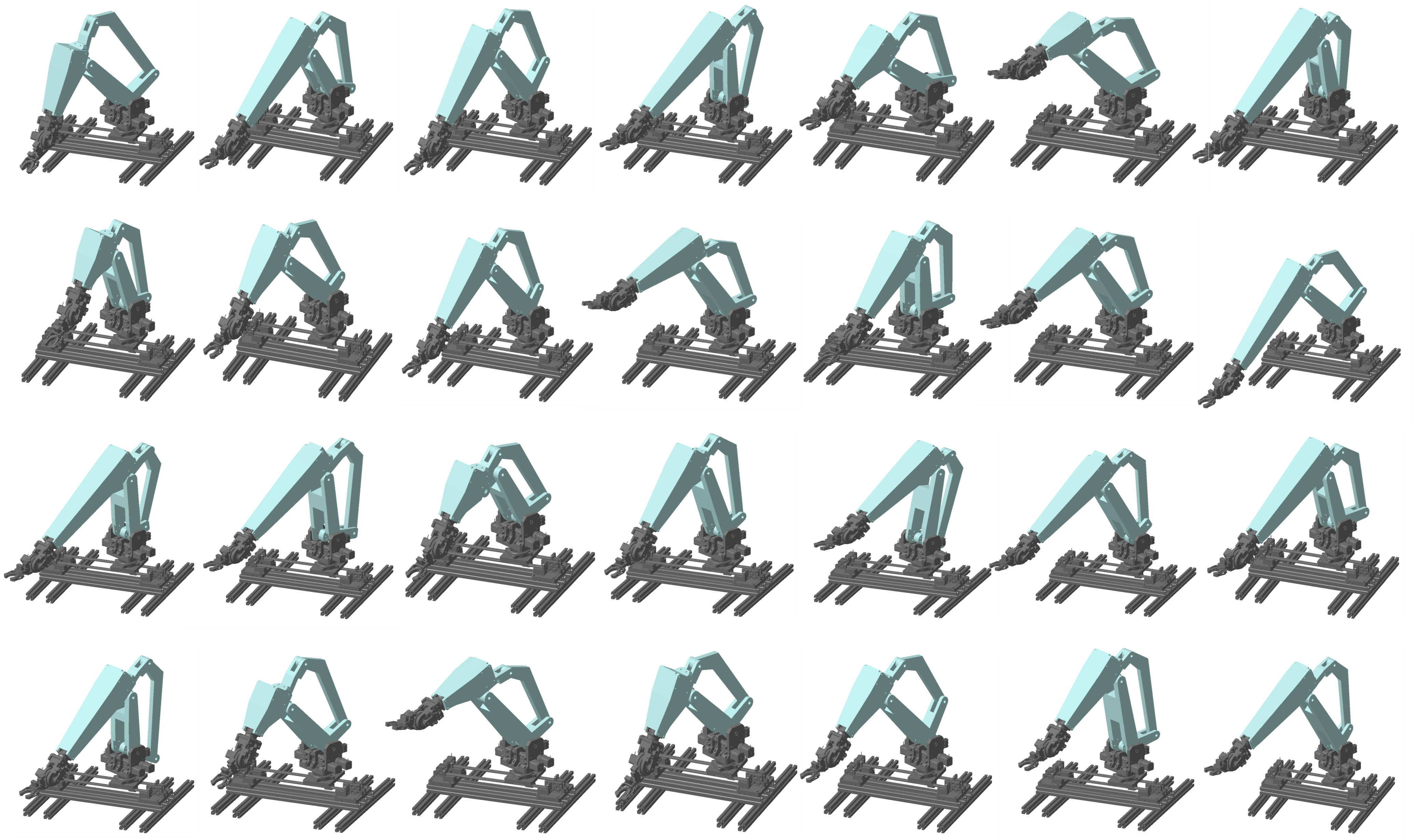}
\captionof{figure}{Parametric variations of the reconstructed robotic-arm assembly. The 28 representative configurations illustrate geometric variations achieved through design parameter adjustments while maintaining assembly relationships and geometric validity. The gray base represents the fixed, non-editable structures.}
\label{fig:robot}
\end{center}

\subsection{Ablation}
\label{sec:ablation}

Removed one of the four input channels at a time and reran only the discrimination stage (4 conditions $\times$ 5 parts $\times$ 5 runs) to examine the effect on the choice of independent variables. Selection count is the number of dimensions selected as independent or dependent variables in a run. Stability is the mean pairwise Jaccard similarity of selection sets across five runs under the same condition. Agreement is the Jaccard similarity between the most frequent selection set under that condition and the most frequent set under the all-channel condition. Lower agreement indicates a larger change in the discrimination outcome after channel removal. Agreement with the all-channel condition in Table~\ref{tab:ablation} measures selection differences from the reference condition, not accuracy against ground truth.

\begin{table}[t]
  \centering
  \caption{Changes in discrimination after channel removal, averaged over five parts. The all-channel condition uses the same runs as the discrimination without requests in Section~4.4.1. Stability is the mean pairwise Jaccard similarity of selection sets across five runs under the same condition. Agreement is the Jaccard similarity between the most frequent selection set under the given condition and that under the all-channel condition.}
  \label{tab:ablation}
  \begin{tabular}{lccc}
    \toprule
    Condition & Selections & Stability & Agreement with all channels \\
    \midrule
    All channels          & 19.3 & 0.67 & 1.00 \\
    $-$ Feature table     & 16.6 & 0.69 & 0.51 \\
    $-$ Geometric constraints & 19.8 & 0.57 & 0.67 \\
    $-$ AAG topology tree & 20.6 & 0.60 & 0.70 \\
    $-$ Rendered images   & 19.2 & 0.66 & 0.83 \\
    \bottomrule
  \end{tabular}
\end{table}

Removing the feature table produced the largest change in the selection outcome. Agreement with the most frequent all-channel set fell to 0.51, and the omitted dimensions were measurements available only from the table, such as inter-axis distances, plane-pair separations, wall thicknesses, and positions. For the joint upper, 12 of the 20 dimensions in the most frequent set disappeared. Stability increased slightly, although this change may partly reflect the reduced candidate information. These results indicate that removing the feature table can affect both the availability of candidate dimension information and the discrimination outcome.

Removing geometric constraints affected consistency more strongly than the selection outcome. Stability showed its largest decline, from 0.67 to 0.57, and outside-graph selections increased. For the rocker, selections varied across the five runs, and agreement between the most frequent sets fell to 0.43. Without the channel specifying which dimensions are coupled, discrimination selected different dimensions as independent variables across runs with the same input.

Removing the AAG topology tree increased the number of selections. This condition produced the highest mean selection count; for the crank, the count increased from 20 to 28.8. The frame's stability was 0.48, the lowest among the four removal conditions. This suggests that removing adjacency information may affect feature-level interpretation of dimensions. Whether individual elements were selected redundantly needs to be checked against the discrimination logs.

Removing rendered images had the smallest effect on discrimination. Although the impact was relatively small in these cases, this result does not generalize to imply that images are unnecessary or redundant for all geometries. Images are also used in the build loop by the verification LLM to compare generated views against the source. Because this experiment covers only discrimination, it does not measure that contribution.

For the tested cases, the feature table determines the available dimensions for selection, geometric constraints support consistent selection, and the AAG groups dimensions at the feature level.

\section{Conclusions and Future Work}
\label{sec:conclusions_future_work}
This study presented a design-intent-aware reverse-parametrization framework for converting history-free STEP models into editable parametric CAD representations. By separating LLM-based interpretation from deterministic geometric construction and verification, the framework establishes design variables and dependencies without task-specific training. The resulting Grasshopper definitions support coordinated component modifications through shared parameters and assembly constraints. Across five mechanical parts and 25 runs, the mean global and analytic-feature-region IoUs were 0.984 and 0.983, respectively, with all runs satisfying the combined shape-preservation criterion. The requested dimension was selected as an independent design variable in 19 of 20 user-request runs. Representative robotic-arm configurations further demonstrated coordinated modifications while maintaining geometric validity, mechanism closure, and joint relations without inter-part interference.

These findings highlight that practical CAD reuse requires meaningful editing controls and explicit dependencies in addition to geometric fidelity. However, generalizability beyond the evaluated components and assembly remains to be established, particularly for irregular and free-form geometries such as topology-optimized shapes. Moreover, sample-based validity checks cannot guarantee feasibility throughout a continuous multidimensional design space. Future work will extend the framework to more complex geometries and assembly structures, investigate alternative parametrization approaches for irregular shapes, and strengthen the validation of coupled parameter changes. These developments would broaden the reuse of history-free CAD models in iterative engineering design and optimization.

\section*{Acknowledgments} 
\paragraph{Funding}
This work was supported by grants from the Ministry of Science and ICT (GTL24033-000, N1026002, and No.~2022-0-00986), the Ministry of Trade, Industry and Energy (RS-2025-02317327 and RS-2025-25444634), the Ministry of Oceans and Fisheries (PET0050), and Korea Hydro \& Nuclear Power Co., Ltd. (No.~8-Tech-07).

\section*{Declarations}
The authors declare that they have no known competing financial interests or personal relationships that could influence the work reported herein.

\section*{Replication of results}
The code and data are available from the corresponding author on reasonable request.


\newpage
\bibliographystyle{cas-model2-names}

\bibliography{cas-refs}


\vskip 3pc

\end{document}